\documentclass[sts]{imsart}

\usepackage{thmtools}
\usepackage{appendix}
\usepackage{bm}
\usepackage{amsfonts}
\usepackage{graphics,epsfig,pstricks}
\usepackage{enumerate}

\usepackage{natbib}

\startlocaldefs

\renewcommand\thmcontinues[1]{Continued}

\newcommand{\nmathbf}{\bm}

\def\bfX{\nmathbf X}

\def\bfx{\nmathbf x}
\def\bfy{\nmathbf y}

\def\bftheta  {\nmathbf \theta}

\def\bfTheta  {\nmathbf \Theta}

\newtheorem{theorem0}{Theorem}
\newtheorem{lemma0}{Lemma}
\newtheorem{remark0}{Remark}
\newtheorem{fact0}{Fact}
\newtheorem{example0}{Example}
\newtheorem{definition0}{Definition}
\newtheorem{corollary0}{Corollary}
\newtheorem{proposition0}{Proposition}
\newtheorem{algorithmY}{Algorithm}

\newenvironment{theorem}{\begin{theorem0} \mbox{} }{\end{theorem0}}
\newenvironment{lemma}{\begin{lemma0} \mbox{}}{\end{lemma0}}
\newenvironment{remark}{\begin{remark0} \mbox{}}{\end{remark0}}

\newenvironment{example}{\begin{example0} }{\end{example0}}
\newenvironment{definition}{\begin{definition0}
\mbox{}}{\end{definition0}}

\newenvironment{proposition}{\begin{proposition0}\mbox{}
}{\end{proposition0}}

\newcommand{\bpro}{\begin{proof}}
\newcommand{\epro}{\end{proof}}
\newcommand{\be}{\begin{eqnarray}}
\newcommand{\ee}{\end{eqnarray}}
\newcommand{\ba}{\begin{eqnarray*}}
\newcommand{\ea}{\end{eqnarray*}}
\newcommand{\bc}{\begin{center}}
\newcommand{\ec}{\end{center}}
\newcommand{\btab}[1]{\begin{tabular}{#1}}
\newcommand{\etab}{\end{tabular}}

\newcommand{\go}{\rightarrow}

\newcommand{\reals}{\mbox{\rm I\kern-.20em R}}
\newcommand{\sreals}{\mbox{\small \rm I\kern-.20em R}}

\newenvironment{proof}{\vspace*{-2mm}\noindent{\sc Proof}. \mbox{} \rm}
   {\hfill\fbox{}}

\endlocaldefs

\begin{document}

\begin{frontmatter}

\title{$p$-Value as the Strength of Evidence Measured by Confidence Distribution}
\runtitle{$p$-Value by Confidence Distribution}


\begin{aug}
\author{\fnms{Sifan} \snm{Liu} \corref{} \ead[label=e1]{sifan.liu@tjufe.edu.cn}}
\author{\fnms{Regina} \snm{Liu} \ead[label=e2]{rliu@stat.rutgers.edu}}
\and
\author{\fnms{Min-ge} \snm{Xie} \ead[label=e3]{mxie@stat.rutgers.edu}}

\thankstext{t1}{Sifan Liu is Instructor, Department of Statistics, Tianjin University of Finance \& Economics, Tianjin, China. Email: sifan.liu@tjufe.edu.cn. Regina Liu and Min-ge Xie are Professors, Department of Statistics, Rutgers University, Piscataway, NJ 08854. Email: rliu@stat.rutgers.edu and mxie@stat.rutgers.edu. 
The authors acknowledge support from grants: NSF DMS1513483, DMS1737857 and DMS-1812048.}

\runauthor{Liu et al.}

\end{aug}

\begin{abstract}
The notion of $p$-value is a fundamental concept in statistical inference and has been widely used for reporting outcomes of hypothesis tests. 
However, $p$-value is often misinterpreted, misused or miscommunicated in practice. Part of the issue is that existing definitions of $p$-value are often derived from constructions under specific settings, and a general definition that directly reflects the evidence of the null hypothesis is not yet available. 
In this article, we first propose a general and rigorous definition of $p$-value that fulfills two performance-based characteristics. 
The performance-based  definition subsumes 
all existing construction-based definitions of the $p$-value, and justifies their interpretations. 
The paper further presents a specific 
 approach based on confidence distribution to formulate and calculate $p$-values.  
This specific way of computing $p$ values  
has two main advantages. 
First, it is applicable for a wide range of hypothesis testing problems, including the standard one- and two-sided tests, tests with interval-type null, intersection-union tests, multivariate tests and so on. 
Second, it can naturally lead to a coherent interpretation of $p$-value as evidence in support of the null hypothesis, as well as a meaningful measure of degree of such support. 
In particular, it places a meaning of a large $p$-value, e.g. $p$-value of $0.8$ has more support than $0.5$. 
Numerical examples are used to illustrate the wide applicability and computational feasibility of our approach.   
We show that our proposal is effective and can be applied broadly, without further consideration of the form/size of the null space. 
As for existing testing methods, the solutions have not been available or cannot be easily obtained.
\end{abstract}


\end{frontmatter}

\section{Introduction}

$P$-value is one of the most popular statistical inference tools. 
It is widely used in decision making process concerning data analysis in many domains.
For example, \cite{pval_report} identified {4,572,043} $p$-values in 1,608,736 MEDLINE abstracts and {3,438,299} $p$-values in 385,393 PubMed Central full-text articles between 1990-2015.
However, $p$-value is frequently misused and misinterpreted in practice. 
For instance, the $p$-value is often misinterpreted either as the probability that the null hypothesis holds, or as the error rate that the null hypothesis is falsely rejected; cf., \cite{Berger2003} and references therein. 
Many concerns have been raised about the practical issues of using $p$-values
(see, e.g., \cite{Nuzzo2014, Baker2016, Was:Lazar:2016, Benjamin2017, Chawla2017}, among many others). 

We speculate that the issues of $p$-value may be partially due to the facts that the traditional $p$-value definitions are not rigorous---the desired features of the performance of $p$-value are not clearly and mathematically presented, and their interpretations are often not straightforward (i.e., $p$-value is {\it not} interpreted as a measure of strength of the evidence obtained from the data {\it in support of} the null hypothesis). 
There is no clarification in the literature on whether a $p$-value provides any evidence for ``accepting'' the null, and the actual meaning of a {\it non-small} $p$-value is always missing.  
In particular, under the same settings, {\it how do we interpret a p-value of 0.80 compared to another one, say, 0.50?}
So far, no precise answer is given. 
This is an important aspect for making inferences in practice, because most people rely on $0.05$ as the threshold to make decisions, but many have also argued that the threshold should be a different value and given by domain experts \citep[cf., e.g.,][]{Adibi2019}.  

The goal of this paper is to provide a broader perspective of $p$-value  that:
\begin{itemize}
\item gives us a more comprehensive understanding of the data, not only restricted to standard hypothesis testing problems (one- and two-sided tests);
\item allows us to extract relevant information from the given dataset in terms of its evidence {\it in support} of a target hypothesis;
\item can be readily used as a decision tool in comparing $p$-value with a given significant level ($\alpha$), when a decision making is needed. 
\end{itemize}
For theoretical justifications, we propose a general and rigorous definition of $p$-value characterized as two performance assessments. 
This formal definition directly relates to the logic behind the $p$-value development and subsumes almost all existing definitions. 
We then propose a concrete approach based on the concept of confidence distribution (CD) (cf., \cite{Xie:Singh:2013} and references therein), to formulate and calculate such $p$-values. The $p$-value calculated by CD satisfies the general performance-based definition. 
We show that this CD-based approach has several advantages:
\begin{itemize}
\item it provides an intuitive and meaningful interpretation of $p$-value as the strength of evidence from the given data in {\it support} of a hypothesis, as well as a meaningful level of degree of such support (e.g. $p$-value of $0.8$ has more support than $0.5$); 
\item it is applicable to a wide range of testing problems beyond the standard one- and two-sided tests, such as tests with odd-shaped null spaces;
\item it enables us to obtain test results directly, without having to explicitly construct a test statistic and evaluate its sampling distribution under the null hypothesis. 
\end{itemize}

\subsection{A brief review of the $p$-value}

While computations of $p$-values date back to the
1770s \citep{Stigler1986}, the $p$-value was first formally introduced by \cite{Pearson1900} in Pearson's chi-squared test for two by two tables.
Start from 1920s, \cite{Fisher:1925, Fisher:1935a, Fisher:1935b} popularized it and made it an influential statistical inference tool. 

{\bf The logic -- } The $p$-value is often used to quantify the statistical significance for ``rejecting'' the targeted statement (null hypothesis).
The logic behind the $p$-value development is \underline{proof by ``contradiction"}:
\begin{itemize}
\item[{\bf [L]}] 
{\it  Assuming the statement is true, the $p$-value is designed as an assessment to evaluate the chance, or how likely, the observed data is ``compatible'' with this statement. If the $p$-value is small, we ``consider'' there is a conflict (contradiction), 
which indicates that the statement fails to account for the whole of the facts
} \citep{Fisher:1925}. 
\end{itemize}
Unlike the usual proof-by-contradiction in (nonrandom) math problems, statistics deal with random phenomena and 
we rarely have $100$\% certainty that a decision (reject or do not reject the statement) is correct.  
To overcome this obstacle,  the frequency (frequentist) argument is often adopted -- rejecting a correct statement should be avoided for {\it majority} of the time.
Here, the actual meaning of ``{\it majority}'' is linked to the chosen threshold value (significance level) which is considered ``{\it small}".
For instance, suppose a statement is correct, we hope that, in $100$ tries, at least $95$ times we can make the correct decisions in not rejecting the statement; then, we choose $5\%$ as the threshold and reject the statement if the $p$-value$\leq 5\%$.

{\bf Textbook definitions -- }
There are two standard ways of defining the $p$-value. Both are tied to a hypothesis testing problem (say, $\mathcal{H}_0: \theta \in \Theta_0$ versus $\mathcal{H}_A: \theta \in \Theta \setminus \Theta_0$) and a test statistic (say, $T(X)$), where $X$ denotes observable data having distribution indexed by $\theta \in \Theta$. Suppose $X=x$ is observed.
The first way defines the $p$-value as an ``upper bound probability'' (cf., e.g., \cite{pval:def}):
\begin{equation} \label{pval_def1}
{pval}_1(x) = \sup_{\theta \in \Theta_0} {P}_{\theta}\{T(X) \geq T(x)\};
\end{equation}
while the second way is based on rejection region $R_\alpha$ of level $\alpha \in (0,1)$ (cf., e.g., \cite{Leh:Rom:2005}):  
\begin{equation} \label{pval_def2}
{pval}_2(x) =  \inf \{ \alpha: T(x) \in R_\alpha\}.
\end{equation}
Both definitions have achieved successes in computing $p$-values in many real applications. 
However, several issues still exist 
(e.g., \cite{Sel:Bay:Berg:2001, Goodman:2008}).
First, as a probability statement in appearance, (\ref{pval_def1}) can easily lead to a widespread misunderstanding that $p$-value is the probability that $\mathcal{H}_0$ is true; while (\ref{pval_def2}), which is based on significant levels (error rates), can cause a common confusion between $p$-value and an error rate. 
Second, 
neither of them provides a {\it direct} connection or a clear evidence-based interpretation to the logic outlined in ${\bf [L]}$.
Specifically, (\ref{pval_def1}) is often interpreted as: assuming $\mathcal{H}_0$ is true, the probability that $T(X)$ is ``as least as extreme (inconsistent with $\mathcal{H}_0$) as" its observed value $t(x)$.
Although logically correct, this bases ${pval}_1(x)$ on non-occurred results somehow inconsistent with $\mathcal{H}_0$, which is {\it indirect} and antagonistic to our interest.
In addition, the connection between (\ref{pval_def2}) and ${\bf [L]}$ is vague and indirect, since the conditioning on $\mathcal{H}_0$ is hidden.
Furthermore,
both definitions require the specific $T(X)$ and/or $R_{\alpha}$ and often limit the constructions of $p$-values to the standard one- or two-sided tests.
These constructions can be complicated or difficult, when $T(X)$ and/or $R_{\alpha}$ are difficult to get, or the distribution of $T(X)$ is not of standard form. 

{\bf Performance-based characteristics -- }
Directly following {\bf [L]},  we notice that \underline{two characteristics} of the performance of the $p$-value are important, and also meet the common understandings in the literature:  
\underline{The first characteristic} is that, {\it when a statement (null hypothesis $\mathcal{H}_0$) is true, the corresponding $p$-value is stochastically equal to or larger than Uniform[0,1]}, which is a formal statement of the consensus that {\it $p$-value typically follows uniform $U[0,1]$ distribution under $\mathcal{H}_0$}  (c.f., e.g., \cite{Berger:Boos:1994,Liu:Singh:1997,Shafer:2011}).  This $U[0,1]$-distributed characteristic is perfectly in line with {\bf [L]} and suggests that if we repeatedly use the defined $p$-value as a tool and reject $\mathcal{H}_0$ when calculated value is smaller than $\alpha \in (0,1)$, the probability of mistakenly rejecting a correct $\mathcal{H}_0$ will be less than 100$\alpha$\%.
\underline{The second characteristic} is that, {\it when a statement is false, the corresponding $p$-value should be getting closer and closer to zero as sample size increases}. 
It can also be rephrased as {\it given significance level $\alpha$, the probability of correctly rejecting a false $\mathcal{H}_0$ (when $p$-value$\leq \alpha$) will be close to one as long as the sample size is sufficient.}
This characteristic ensures that we will be able to tell apart ${\Theta}_0$ and its complement $\Theta \setminus \Theta_0$ as more and more information is collected. 
Indeed, 
the two characteristics above insure the performance of a test using $p$-value, in controlling Type-I error under the null and ensuring testing power under the alternative.

In the literature, there are several different definitions and interpretations of $p$-values other than the textbook versions. 
For example, \cite{Scher:1996} discussed on a unified version of the $p$-value for one-sided and two-sided tests in certain scenarios. 
\cite{Mud:Cha:2009} introduced a generalized $p$-value definition which depends on a partial ordering of random variables and constructed $p$-values using the results under Neyman-Pearson framework. 
\cite{Martin:Liu:2014} gave an interpretation by plausibility function under the framework of inferential model. 
See also \cite{Bickel:Doksum:1977, Tsui1989, Couso:San:2008, Patriota:2013} for other developments from different approaches.
However, in all above cases, neither an evidence-based interpretation of the $p$-value as the strength of evidence in support of the statement, nor a unified and rigorous formulation of the $p$-value, is given. 


\subsection{Arrangements of the paper}

In Section 2, we propose a formal and performance-based definition of the $p$-value, directly linked with the key logic [{\bf L}].
Our proposal subsumes the textbook definitions as well as the so-called limiting $p$-value defined in (but not limited to) the bootstrap literature \citep{Beran:1986,Singh:Berk:1994, Liu:Singh:1997}. 
Based on this definition, we are able to broaden the concept of $p$-value to a mapping that assesses the strength of evidence obtained from data supporting a statement.
In Section 3, we propose a concrete approach using the \underline{confidence distribution (CD)} (cf., e.g.,  \cite{Xie:Singh:2013, Schweder:Hjort:2016}).
Under CD, we formulate and interpret the $p$-value as {\it a support of the null space $\Theta_0$}. 
Specifically, in Section 3.1, we first give a brief review of the CD concept, and then introduce {\it direct support} and {\it indirect support} under CD, which provides an evidence-based interpretation of $p$-value for the standard one- and two- sided test, respectively. 
Furthermore, to pursue a potential unification, we propose {\it full support} by combining direct and indirect support.
In Section 3.2, we present a unified construction of $p$-value for univariate cases, based on the supports under CD.
We show that our proposal satisfies the performance-based definition, and typically agrees with the textbook $p$-values. 
More importantly, we show that the proposal is also applicable for a wide range of hypothesis testing problems including: 
i) tests with interval null hypotheses, which are motivated by many practical problems \citep{Hodges:Lehmann:1954, Balci:Sargent:1981, Balci:Sargent:1982, Freedman1984};
ii) the intersection-union test, of which a special case is the widely-used bio-equivalence test (c.f. \cite{Schuirmann:1981,Schuirmann:1987,Anderson:Hauck:1983,Berger:Hsu:1996}).
In Section 3.3, we discuss on the general guidelines of our CD-based construction of $p$-value mappings. 
In Section 4, we extend our proposal to tackle with multivariate hypothesis testing problems, where the form/shape of the null space can be various. 
In such cases, we formulate the supports based on the limiting $p$-values given by \cite{Liu:Singh:1997} using data depth (c.f., e.g., \cite{Liu:1990}) and bootstrap. 
Numerical examples are conducted in Section 5 to illustrate the broad applicability and computational feasibility of this approach.
We show that our proposal is a safe and universally effective approach one can be applied broadly, without further consideration of the form/size of the null space. 
Especially, other than standard tests, we consider the situations where the null space is a small interval, a union of small intervals or a small region. 
As for existing testing methods, the solutions have not been available or cannot be easily obtained.

\section{A GENERAL Definition of $p$-Value Based on Performance}
Let $\bfX_n$ denote the random sample of size $n$ from a distribution indexed by a parameter $\theta \in \Theta$. 
Let $\mathcal{B}_{\Theta}$ be the Borel algebra of the parameter space $\Theta$, and $\mathbb{X}^n$ be the sample space corresponding to observed sample data $\bfx_n$. 
Consider the statement of interest $\mathcal{H}_0: \theta \in \Theta_0$, where $\Theta_0 \in \mathcal{B}_{\Theta}$. Let $p(\cdot, \cdot)$ be a mapping: $\mathbb{X}^n \times \mathcal{B}_{\Theta} \mapsto [0,1]$. 
We propose a {\it performance-based} definition of $p$-value as follows. 
\begin{definition}\label{pval_def_map}
(A) The value of $p(\bfx_n,\Theta_0)$ is called a \underline{$p$-value} for the statement $\mathcal{H}_0 : \theta \in \Theta_0$, if $p(\bfX_n, \Theta_0)$, as a function of the random sample $\bfX_n$, satisfies the following conditions for any $\alpha \in (0,1)$,

(i) ${P}_{\theta} \{ p(\bfX_n, \Theta_0) \leq \alpha \} \leq \alpha$, ~for~all~ $\theta \in { \Theta}_0$; 

(ii) ${P}_{\theta} \{ p(\bfX_n, \Theta_0) \leq \alpha \} \go 1$, as $n \go \infty$, ~for~all~ $\theta \in \Theta \setminus {\Theta}_0$.

\noindent
(B) The value of $p(\bfx_n, \Theta_0)$ is called a \underline{limiting $p$-value ($LP$)} for $\mathcal{H}_0$, if condition~(i) is replaced by the following asymptotic condition:

(i') $\limsup_{n \go \infty} {P}_{\theta} \{ p(\bfX_n, \Theta_0) \leq \alpha \} \leq \alpha$, for all $\theta \in { \Theta_0}$.
\end{definition}

The conditions (i) and (ii) above highlight the performance-based characteristics of $p$-value directly linked to the key logic {\bf [L]}. 
Given a significance level $\alpha \in (0,1)$, we require that: 1) the probability of mistakenly rejecting a correct $\mathcal{H}_0$ be at most $100\alpha\%$; 2) the probability of correctly rejecting a false $\mathcal{H}_0$ be getting closer and closer to one as sample size increases.
Consider the hypothesis testing problem with the null hypothesis $\mathcal{H}_0$ versus the alternative (say, $\mathcal{H}_A: \theta \in \Theta \setminus \Theta_0$),
conditions (i) and (ii)  specify the performance of a test by controlling Type-I error under $\mathcal{H}_0$ and ensuring power under $\mathcal{H}_A$, respectively. 


Typically, to show that $p(\bfX_n,\Theta_0)$ is a $p$-value mapping, one needs to show that $p(\bfX_n,\Theta_0)$ is stochastically equal to or larger than Uniform[0,1] for all $\theta \in \Theta_0$, and degenerates to 0 for all $\theta \in \Theta \setminus \Theta_0$.
The following proposition indicates that the textbook approaches (\ref{pval_def1}) and (\ref{pval_def2}) provide such mappings, as consequences, the textbook $p$-values satisfy {\it Definition \ref{pval_def_map}}. A rigorous proof is given in Appendix A. 
\nocite{Bahadur1971}
\begin{proposition} \label{Prop1}
Suppose $T: \mathbb{X}^n \mapsto \mathbb{R}$ is a test statistic constructed so that a large value of $T(\bfX_n)$ contradicts $\mathcal{H}_0$, and that for any $\theta \in \Theta_0$, the exact cumulative distribution function (c.d.f.) of $T(\bfX_n)$ (denoted by $G_{T,\theta}$) exists.
Then, ${pval}_1$ in (\ref{pval_def1}) and ${pval}_2$ in (\ref{pval_def2}) satisfy {\it Definition \ref{pval_def_map}}.
\end{proposition}

Many $p$-values in real applications are derived from the limiting null distributions of test statistics (i.e., $G_{T,\theta}$ is asymptotical), and they are approximations of the ``exact'' $p$-values. 
These approximations are often {\it limiting p-values (LPs)}, which are also defined in {\it Definition \ref{pval_def_map}}.
In the following, Example \ref{eg1} presents a frequently-used $LP$ for testing about a normal mean, and Example \ref{eg_LP} (c.f., e.g., \cite{Liu:Singh:1997}) discusses a more general situation where the exact computation of the $p$-value is extremely difficult. 

\begin{example}\label{eg1} Consider a sample data ${\bfy}_n = (y_1, \ldots, y_n)$, from $N(\theta, \sigma^2)$ with both $\theta$ and $\sigma^2$ unknown. 
Then for the left one-sided test 
\be \label{left_one_sided}
\mathcal{H}_0 : \theta \leq \theta_0 {~versus~} \mathcal{H}_A: \theta > \theta_0,
\ee
a $LP$ based on z-test is 
\be \label{eg_pval_LP}
p({\bfy}_n, (-\infty, \theta_0])
 = \Phi(\sqrt{n}(\theta_0  -{\bar y}_n)/s_n),
\ee
where $\Phi$ is the c.d.f. of standard normal, ${\bar y}_n$ is the sample mean and $s_n$ is the sample standard deviation. 

Instead, let $F_{t_{n-1}}$ be the c.d.f. of the $t_{n -1}$-distribution, we have 
$
p({\bfy}_n, (-\infty, \theta_0])
 = F_{t_{n-1}}(\sqrt{n}(\theta_0  -{\bar y}_n)/s_n),
$
which is the (exact) p-value obtained by the classical $t$-test. 
\end{example}

\begin{example}\label{eg_LP} 
Consider testing the population mean $\theta$ in the left one-sided test (\ref{left_one_sided}) as in Example \ref{eg1}. Here, we concentrate on distributions with finite variances: $\mathcal{H}_0: \{ F \mid \int y dF \leq \theta_0, \int y^2 dF < \infty  \}$ versus $\mathcal{H}_A: \{ F \mid \int y dF > \theta_0, \int y^2 dF < \infty  \}$. 
Then, given sample data $\bfy_n$, (\ref{eg_pval_LP}) is still a $LP$ by the central limit theorem.  
\end{example}

Although it does not provide a specific way to construct $p$-values, {\it Definition \ref{pval_def_map}} provides two fundamental requirements that ensure the argument of ``proof by contradiction'' in [{\bf L}] can strictly go through mathematically. 
In particular, (i) or (i') is a basic requirement for a $p$-value, since controlling the size of a test is of primary concern for designing the test; 
while (ii) is a minimal requirement for $p$-values on the power of the test, and one could seek testing procedure with appropriate mapping for achieving better power or other purposes.  
Through mappings over the sample space and the parameter space, {\it Definition \ref{pval_def_map}} can cover almost all $p$-values available in  statistics literature. 
It enables us to justify any candidate of $p$-value mapping, and guarantees the desired features of using the defined $p$-values. 
More importantly, it allows us to broaden the concept of $p$-value to a mapping measuring the strength of evidence coming from the observations $\bfx_n$ in support of the null space $\Theta_0$. 

Note that in {\it Definition \ref{pval_def_map}}, if $\Theta_0$ is not a closed set, $\Theta_0$ in the required conditions may be replaced by its closure set ${\overline \Theta}_0$ (the set contains all the boundary limit points). Since this situation is very rare in real applications, we shall assume $\Theta_0$ is closed throughout this article.  
In addition, it is not difficult to see our proposed $p$-values tend to zero under the corresponding alternative hypotheses, since the $p$-value mappings are not properly centered and will vanish outside of $\Theta_0$ with large $n$. 
We shall avoid repeating this observation and focus on condition (i) or (i') in justifications.

\section{$p$-Values Based on Confidence Distribution}

Our proposed performance-based definition is rigorous, but it does not provide a specific way to construct $p$-values. 
Whenever applicable, we can use the textbook approaches to compute $p$-values. In Section 3 \& 4, we propose an alternative approach that uses confidence distribution (CD) to formulate and calculate $p$-values.
The benefits of this CD-based construction include:
\begin{itemize}
\item for a wide range of hypothesis testing problems, it satisfies {\it Definition 1};
\item through CD supports, it affords an interpretation of the $p$-value as the strength of evidence in support of the null.  
\end{itemize}

In the following, we first review the concept of CD and then propose CD-based notions of $p$-value for univariate hypothesis testing problems. Multivariate cases will be discussed in Section 4.  

\subsection{The Concept of CD Supports}
\subsubsection{\textsf{A brief review of CD and its connection to $p$-value}} \mbox{}

From the estimation's point of view, CD is a ``distribution estimator" of the parameter of interest in frequentist inference. 
CDs are to provide ``simple and interpretable summaries of what can reasonably be learned from data (and an assumed model)" \citep{Cox2013}. 
A formal definition of CD (cf., e.g.,  \cite{Xie:Singh:2013, Schweder:Hjort:2016}) is as follows: 

\begin{definition} \label{def_CD}
A function $H_{n}(\cdot) = H(\bfx_n, \cdot)$ on $\mathcal{X} \times \Theta \go [0,1]$ is called a {\underline{confidence distribution (CD)}} for a parameter $\theta$, if it follows two requirements: (i) for each given sample set $\bfx_n \in \mathcal{X}$, $H_{n}(\cdot)$ is a continuous cumulative distribution function on $\Theta$; (ii) the function can provide confidence intervals (regions) of all levels for $\theta$.
\end{definition}

Here, $(i)$ emphasizes the distribution-estimator nature of CD, while $(ii)$ is imposed to ensure that the statistical inferences derived from the CD have desired frequentist properties linked to confidence intervals (CI). 
When $\theta$ is univariate, $(ii)$ indicates that {\it at the true parameter value $\theta=\theta_0$, $H_{n}(\theta_0) = H(\bfx_n, \theta_0)$, as a function of the sample set $\bfx_n$, follows ${Uniform}[0, 1]$.}

\begin{example}\label{eg_simple} Consider the settings in Example \ref{eg1}. For simplicity, assume $\sigma^2=1$. 
Immediately, we have a point estimate of $\theta$ as ${\bar y}_n$, an interval estimate (95\% CI) as $({\bar y}_n-1.96/\sqrt{n}, {\bar y}_n+1.96/\sqrt{n})$ and a sample dependent distribution function on $\Theta$ as $N({\bar y}_n,1/n)$, of which the c.d.f. is $H_n(\theta)=\Phi(\sqrt{n} (\theta-{\bar y}_n))$. 

Here, $H_n$ is a CD of $\theta$. Notice that $H_n$ can provide CIs of all levels. For $\alpha \in (0,1)$, a 100($1-\alpha$)\% CI of $\theta$ is $(H_n^{-1}(\alpha/2), H_n^{-1}(1-\alpha/2))=({\bar y}_n-\Phi^{-1}(\alpha/2)/\sqrt{n}, {\bar y}_n+\Phi^{-1}(1-\alpha/2)/\sqrt{n})$.
Also, the mean (median) of $N({\bar y}_n,1/n)$ is the point estimator, and the tail mass $H_n(\theta_0)$ is a $p$-value for testing (\ref{left_one_sided}). 

Therefore, as a typical CD of $\theta$, $N({\bar y}_n,1/n)$ provides meaningful information for making inferences about $\theta$. 
Note that $H_n$ also matches the Bayesian posterior of $\theta$ with a flat prior.
\end{example}

If the requirement $(ii)$ is true only asymptotically and the continuity requirement on $H_n$ is dropped, the function $H_n(\cdot)$ is called an {\it asymptotic CD (aCD)} \citep{Xie:Singh:2013}.

\begin{example}[continues=eg1]
An aCD of $\theta$ can be obtained based on normal approximation, 
$H_{n}^{A} (\theta) = \Phi \left( \sqrt{n}[\theta-\bar{y}_n]/s_n \right)$, which matches the form in (\ref{eg_pval_LP}) exactly. 
\end{example}


Although CD is a purely frequentist concept, it links to both Bayesian and fiducial inference concepts. 
``Any approach, regardless of being frequentist, fiducial or Bayesian, can potentially be unified under the concept of CDs, as long as it can be used to build confidence intervals (regions) of all levels, exactly or asymptotically'' \citep{Xie:Singh:2013}. 
Some examples of CDs include: bootstrap distributions \citep{Efron:1982}, $p$-value functions \citep{Fraser:1991}, Bayesian posteriors, normalized likelihood functions, etc.  

Particularly, to illustrate the connection between CD and $p$-value function, consider common situations where there exists a pivot $U\{T(\bfX_n), \theta\}$ with continuous c.d.f. $G_U$, independent from $\bfX_n$ and $\theta$.  
Suppose $U\{T(\bfX_n), \theta\}$ is increasingly monotonic with respect to $T(\bfX_n)$ and has the form
$
U=({{\hat \theta}-\theta})/{SE({\hat \theta})},
$
where ${\hat \theta}$ is an arbitrary estimator and $SE({\hat \theta})$ is the standard error. 
For the left one-sided test (\ref{left_one_sided}), we construct a mapping based on (\ref{pval_def1}):
\small
\be \label{expression}
~~~~{pval}_1(\bfx_n)=\sup_{\theta \in ( -\infty, \theta_0]}P_{U} (U \geq u)=P_{U} \left( U \geq \frac{{\hat \theta-\theta_0}}{SE({\hat \theta})} \right)=1-G_U\left( \frac{{\hat \theta-\theta_0}}{SE({\hat \theta})} \right),
\ee
\normalsize
where $u$ denotes the observed value of $U$, and correspondingly, ${\hat \theta}$ is the sample-dependent estimate.
Given $\alpha \in (0,1)$, $\{ {pval}_1(\bfx_n) \leq \alpha \}$ can be written as
\small
\ba
\left\{ G_U\left( \frac{{\hat \theta-\theta_0}}{SE({\hat \theta})} \right) \geq 1-\alpha \right\} 
= \left\{  \frac{{\hat \theta-\theta_0}}{SE({\hat \theta})} \geq q_{1-\alpha} \right\}
= \left\{  \frac{{\hat \theta-\theta}}{SE({\hat \theta})} \geq q_{1-\alpha} +  \frac{{\theta_0-\theta}}{SE({\hat \theta})} \right\},
\ea
\normalsize
where $q_{1-\alpha}$ is the $1-\alpha$ quantile of $G_U$. 
Clearly, (\ref{expression}) satisfies both (i)  and (ii) in {\it Definition \ref{pval_def_map}}. Note that when $n \go \infty$, $SE({\hat \theta})$ tend to zero for any reasonable estimator $\hat \theta$. 
Immediately, 
we have the following observations:
$(i)$ given $\bfx_n \in \mathbb{X}^n$, let $\theta_0$ vary in $\Theta$, ${pval}_{1}(\bfx_n)$ is a c.d.f. on $\Theta$;
$(ii)$ let $\theta_0$ be the true value of $\theta$ and $\bfx_n$ be random, ${pval}_{1}(\bfx_n)$ as a function of $\bfx_n$ follows {Uniform}$[0,1]$.
Therefore, ${pval}_{1}(\bfx_n)$ is a typical CD of $\theta$.

\subsubsection{\textsf{Direct support under CD}} \mbox{}

Let $H_n(\cdot) = H_n(\cdot; {\bfx}_n)$ denote a CD of $\theta$, and $dH_n$ be the corresponding density function. 
For a subset on the parameter space, we consider a measure of our ``confidence" that the subset covers the true value of $\theta$. 

\begin{definition}
Let  $\Theta_0 \subseteq \Theta$. The \underline{direct support (or evidence)} of $\Theta_0$ under CD $H_n(\theta)$ is
\begin{equation}
\label{direct:support}
 S_{n}^D(\Theta_0) = \int_{\theta \in \Theta_0} dH_n(\theta).
\end{equation}
\end{definition}
 
The direct support, also called ``strong support'' \citep{Sin:Xie:Str:2007}, is a typical ``measure of support'' (cf., e.g., \cite{Scher:1996}).
The motivation here is to look at the CD of $\theta$ as a plausible distribution to which $\theta$ belongs, conditioning on the given data.
Intuitively, the higher the support of $\Theta_0$ is, the more likely an estimate of $\theta$ falls inside $\Theta_0$, thus it is more plausible that $\theta \in \Theta_0$.  
As a special case, if a Bayesian posterior is used as a CD, (\ref{direct:support}) is equivalent to the posterior probability of $\theta \in \Theta_0$.

Based on the previous discussions, (\ref{expression}) is the textbook $p$-value for the one-sided test with $\Theta_0=(-\infty, \theta_0]$. In the meanwhile, (\ref{expression}) provides a $p$-value mapping, leading to the fact that the direct support  
$S_{n}^D((-\infty, \theta_0])=H_n(\theta_0)$. Clearly, the argument is still valid for $\Theta_0=[\theta_0, \infty)$. Then,  {the connection between direct support and {\bf one-sided $p$-value} can be summarized as follows: 
\begin{proposition}
If $\Theta_0$ is of the type $(-\infty, \theta_0]$ or $[\theta_0, \infty)$, the textbook p-value typically agrees with the direct support $S_{n}^D(\Theta_0)$. 
\end{proposition}

The following lemma illustrates the properties of the direct support, which applies to not only the one-sided tests, but also a wider set of problems---a union of intervals.
Here, we assume the following regularity condition: as $n \go \infty$, for any finite $\theta_1$, $\theta_2$ and positive $\epsilon$, $\delta$,
\ba
\sup_{\theta \in [\theta_1,\theta_2]} {P}_{\theta} (\max\{ H_{n}(\theta-\epsilon), 1- H_{n}(\theta+\epsilon) \} > \delta) \go 0.
\ea
The proof of Lemma 1 is given in Appendix B.

\begin{lemma}\label{lemma1}
(a) Let $\Theta_0$ be of the type $(-\infty,a]$ or $[b,\infty)$. Then, for any $\alpha \in (0,1)$, $\sup_{\theta \in \Theta_0} {P}_\theta \left( {S}_{n}^D(\Theta_0) \leq \alpha \right) = \alpha$.

(b) Let $\Theta_0=\cup_{j=1}^k \Theta_{0j}$ where $\Theta_{0j}$ are disjointed of the type $(-\infty,a]$, $[b,\infty)$ or $[c,d]$. Here, $c<d$. If the regularity condition holds, then 
$\sup_{\theta \in \Theta_0} {P}_{\theta} ({S}_{n}^D(\Theta_0) \leq \alpha) \go \alpha$, as $n \go \infty$, for any $\alpha \in (0,1)$.
\end{lemma}

In the above cases, the $p$-value can be calculated and interpreted as the direct support of the null space, which seals the fact that {\it $p$-value is used to measure the strength of evidence ``supporting'' the null}. 
This is a factual argument in comparison with the widespread but indirect statement that $p$-value measures evidence ``against'' the null (cf., e.g, \cite{Mayo:Cox:2006}). 
In the meanwhile, (\ref{direct:support}) provides a CD-based measure of the degree/strength of the support. 
To encompass the ``measure of support'' properties and $p$-value's evidence-based interpretation, our approach places a meaning of large $p$-value, e.g. {\it $p$-value of 0.8 has more support than 0.5}. 
This is very similar to the Bayesian posterior probability. 
It is well-known in Bayesian perspective that, if we choose non-informative priors for location parameters: 1) Bayesian credible intervals match the corresponding confidence intervals guaranteeing the frequentist coverage; 2) the posterior probabilities of the null hypothesis typically agree with $p$-values for the one-sided tests.  
Thus, this ``coincidence'' between CD and Bayesian inferences is a clarification rather than a misinterpretation.

\begin{remark} \label{remark_Direct}
Although $S_n^D(\Theta_0)$ calculates and interprets the corresponding $p$-value in a wide range of problems, its usefulness is limited. Since the CD density is generally continuous, when $\Theta_0$ is narrow, $S_{n}^D(\Theta_0)$ would almost always be small unless $n$ is sufficiently large (c.f., Figure \ref{fig.h0} \& \ref{fig.h3} in the simulation study); more extremely, when $\Theta_0$ is degenerated to a singleton, $S_{n}^D(\Theta_0)$ would be simply zero regardless of where $\Theta_0$ lies. 
In such cases, due to the fact that the width of $\Theta_0$ has a non-negligible effect on the value of $S_{n}^D(\Theta_0)$, an alternative way of evaluating the evidence to $\mathcal{H}_0$ is desired.
\end{remark}

\subsubsection{\textsf{Indirect support under CD}} \mbox{}

To avoid the undesired influence of the width of $\Theta_0$ in measuring its support, we propose another measure of the strength of evidence, called {indirect support}, as follows.
\begin{definition}
The \underline{indirect support (or evidence)} of a subset $\Theta_0 \subset \Theta$ under CD $H_n$ is 
\begin{eqnarray} \label{indirect:support}
{S}_{n}^{IND}( \Theta_0 ) = \inf_{\theta_0 \in \Theta_0} 2 \min \{ H_n(\theta_0), 1-H_n(\theta_0) \}.
\end{eqnarray}
\end{definition}
Clearly, when $\Theta_0$ is a singleton, say $\{ \theta_0 \}$, we have 
\begin{eqnarray} \label{confidence_curve}
{S}_{n}^{IND}( \theta_0 ) \equiv {S}_{n}^{IND}( \{ \theta_0 \} ) = 2 \min \{ H_n(\theta_0), 1-H_n(\theta_0) \}. 
\end{eqnarray}

The motivation here is to exam how plausible it is to assume $\theta =\theta_0$. To facilitate such an examination on some $\{ \theta_0 \}$, for which $S_n^D(\theta_0)$ does not work, we build up some room to consider the opposites of $\theta_0$. Denote $\Theta_{lo}=(-\infty, \theta_0)$, $\Theta_{up}=(\theta_0,\infty)$, respectively. Then, we have 
\ba
{S}_{n}^{IND}( \theta_0 ) 
&=& 2 \min \{1-{S}_{n}^D(\Theta_{up}),1-{S}_{n}^D(\Theta_{lo})\}
\\
&=& 2 [1-\max \{ {S}_{n}^D(\Theta_{up}),{S}_{n}^D(\Theta_{lo})\}]. 
\ea 
Like the proverb said, ``the enemy of my enemy is my friend". In the form of indirect support, we first consider the direct support of ``enemies'' of $\{ \theta_0 \}$, $S_n^D(\Theta_{up})$ and $S_n^D(\Theta_{lo})$. Next, by max$\{ {S}_{n}^D(\Theta_{up}), {S}_{n}^D(\Theta_{lo}) \}$, we take the stronger side (``the tougher enemy'') as a measure of evidence ``against'' $\{ \theta_0 \}$. Then, ``the enemy of enemy", $1-${max}$\{  {S}_{n}^D(\Theta_{up}), {S}_{n}^D(\Theta_{lo}) \}$, can be used to measure the indirect evidence to $\Theta_0$ on one direction (side). Finally, to adjust two directions (sides), we multiply the value above by $2$. 
Intuitively, when $\theta_0$ is extreme in either direction, (\ref{confidence_curve}) will be small, indicating that $\theta_0$ is implausible.


The following proposition implies that the above interpretation through the indirect support can be applied for {\bf two-sided $p$-values}.

\begin{proposition}
If $\Theta_0$ is a singleton $\{ \theta_0 \}$, the textbook $p$-value typically agrees with ${S}_{n}^{IND}(\theta_0)$.
\end{proposition}

The justification of this proposition is given in Lemma 2 ($\gamma=0.5$). 
Briefly speaking, if $\theta_0$ is the true value, we have the key facts that $H_n(\theta_0) \sim  {Uniform}[0,1]$ and $1-H_n(\theta_0) \sim  {Uniform}[0,1]$. Then, ${S}_{n}^{IND}( \theta_0 ) = 2 \min \{ H_n(\theta_0), 1-H_n(\theta_0) \} \sim {Uniform}[0,1]$.
\begin{lemma}
 Let $\Theta_0$ be a singleton $\{ \theta_0 \}$,  ${P}_{\theta=\theta_0} \left( \min \left\{ \frac{H_{n}(\theta_0)}{\gamma},  \frac{1-H_{n}(\theta_0)}{1-\gamma} \right\} \leq \alpha \right)=\alpha$, where $\gamma \in (0,1)$.
\end{lemma}

More generally, when $\Theta_0$ is a subset, we evaluate all the points $\theta_0 \in \Theta_0$ by $S_n^{IND}$ and report the smallest value of support. A large value of $S_n^{IND}(\Theta_0)$ indicates no extreme (inconsistent) value is contained in $\Theta_0$, implying that $\Theta_0$ is plausible; while a small value means that $\Theta_0$ contains some extreme (inconsistent) values, and $\Theta_0$ is plausible only if the direct support of it is large. When $\Theta_0$ is one-tailed interval, $S_n^{IND}(\Theta_0)$ is simply zero.  Therefore, the indirect support can be treated as a useful and necessary complement of the direct support. It is then intuitive that we may use a combination of the direct and the indirect support to measure the strength of evidence. 
Later on, we will propose to combine these two types of support to form a unified notion of $p$-values for both one- and two-sided tests. 


\subsubsection{Full support under CD} \mbox{}

Based on the previous discussions, direct and indirect supports can provide evidence-based interpretations of one- and two-sided $p$-values, respectively. However, the one- and two-sided $p$-values are treated very differently in terms of both calculation and interpretation. We propose a combined measure of evidence, which can fill this gap.
\begin{definition}
Let  $\Theta_0 \subseteq \Theta$. The \underline{full support (or evidence)} of $\Theta_0$ under CD $H_n(\theta)$ is
\begin{equation} \label{full:support}
S^+_n(\Theta_0) = {S}_{n}^D(\Theta_0) +  {S}_{n}^{IND} (\Theta_0).
\end{equation}
\end{definition}

Here, $S^+_n(\Theta_0)$ has two parts: measures of the direct and indirect evidence in support of $\Theta_0$. The former is the distribution estimated measurement of $\Theta_0$, while the latter measures an adjustment based on indirect evidence to ``the enemy of enemy" for $\Theta_0$. Altogether, $p$-values are computed based on a combination of the direct and indirect parts.   

Consider the {\bf conventional one-sided and two-sided} hypothesis tests.
First, for one-sided tests with $\Theta_0=(-\infty, a]$ or $[b,\infty)$, (\ref{full:support}) is
$
S^+_n(\Theta_0) =S_{n}^D(\Theta_0) + 0 =S_{n}^D(\Theta_0), 
$
i.e., the direct support. 
Second, for two-sided tests with $\Theta_0=\{ \theta_0 \}$, 
$
S^+_n(\Theta_0) = 0 + {S}_{n}^{IND}(\theta_0) ={S}_{n}^{IND}(\theta_0),
$
i.e., the indirect support.
Therefore, for both one- and two-sided tests, (\ref{full:support}) matches the textbook $p$-value. 
Furthermore, the definition of (\ref{full:support}) is very general and can accommodate a wide range of testing problems.

\noindent
{\bf The validity of $S^+({\bf X}_n, \Theta_0)$ when $\Theta_0$ is an interval --}
Consider the situations where $\Theta_0$ belongs to the following set of intervals,
\be \label{interval_type}
\mathcal{A} = \{ [c,d]: c,d \in \overline{\mathbb{R}} ~{and}~ c \leq d \},
\ee
where $\overline{\mathbb{R}}$=$\mathbb{R} \cup \{-\infty, +\infty \}$ is the extended real number system. It is clear that the null spaces in one- and two-sided tests are special cases. 

To justify the validity, we first introduce a lemma on combining a $p$-value mapping with a ``degenerated'' mapping. The proof is given in 
Appendix C. 
\begin{lemma} \label{lemma_combination}
Suppose that, $p_1(\bfX_n,\Theta_0)$ is a $p$-value (or $LP$) of the statement $H_{0}: \theta \in \Theta_{0}$. 
Let $q(\bfX_n, \Theta_0)$ be a mapping $\mathcal{X}^n \times \mathcal{B}_{\Theta} \mapsto [0,1]$ satisfying that 
\ba
{P}_\theta \{ q(\bfX_n, \Theta_0) \leq \alpha \} \go 1, ~as~ n \go \infty, ~for~all~ \theta \in \Theta \setminus {\Theta}_0, ~for~any~ \alpha \in (0,1).
\ea
Then, $p_2(\bfX_n, \Theta_0)= p_1(\bfX_n, \Theta_0) + q(\bfX_n, \Theta_0)$ is another $p$-value (or $LP$) of $H_{0}$. 
\end{lemma}

Note that unless $\Theta_0$ is degenerated, $\inf_{u \in \Theta_0} {S}_{n}^{IND} (u) \go 0$, as $n \go \infty$, for $\theta \in \Theta_0$. 
In such cases, $S^+_n(\Theta_0)$ is a $p$-value (or $LP$), because of the nice properties of ${S}_{n}^D (\Theta_0)$. Based on Lemma 1, 2 \& 3, we summarize the resulting features as follows.
\begin{theorem}
Consider the mapping $S^+_n(\Theta_0)$ defined in (\ref{full:support}).
(a) Let $\Theta_0$ be of the type $(-\infty,a]$ or $[b,\infty)$. Then $\sup_{\theta \in \Theta_0} {P}_\theta \left\{ S^+_n(\Theta_0) \leq \alpha \right\} = \alpha$.

(b) Let $\Theta_0$ be a singleton $\{ \theta_0 \}$, ${P}_{\theta=\theta_0} \left\{ S^+_n(\{ \theta_0 \}) \leq \alpha \right\} = \alpha$.

(c) Let $\Theta_0 \in \mathcal{A}$. If the regularity condition holds, 
$\sup_{\theta \in \Theta_0} {P}_{\theta} (S^+_n(\Theta_0) \leq \alpha) \go \alpha, ~as~ n \go \infty.$

(d) Let $\Theta_0=\cup_{j=1}^k \Theta_{0j}$ where $\Theta_{0j}$ are disjointed of the type $(-\infty,a]$, $[b,\infty)$ or $[c,d]$. Here, $c<d$. If the regularity condition holds, then 
$\sup_{\theta \in \Theta_0} {P}_{\theta} ({S}_{n}^+(\Theta_0) \leq \alpha) \go \alpha$, as $n \go \infty$, for any $\alpha \in (0,1)$.
\end{theorem}

\begin{example}[continues=eg1] 
Consider $H_0:\theta \in [a,b]$ vs. $H_A:\theta \in  (-\infty,a)\cup(b,\infty)$. A $LP$ is 
\ba
S^+_n(\Theta_0) &=& S_{n}^D([a,b]) + 2\min \{ S_{n}^D((-\infty,a)), S_{n}^D((b, \infty)) \} \\
&=&   
\left\{ 
\begin{array}{l l}
\Phi(\sqrt{n}({\bar y}_n-a)/s_n) + \Phi(\sqrt{n}({\bar y}_n-b)/s_n)
 & {~if~{\bar y}_n < (a+b)/2}; \\
\Phi(\sqrt{n}(a-{\bar y}_n)/s_n) + \Phi(\sqrt{n}(b-{\bar y}_n)/s_n)
  & {~if~{\bar y}_n \geq (a+b)/2}.
\end{array}
 \right.
\ea
This result is an ``n-sample" version of the $p$-value given by \cite{Scher:1996}, which is derived from the corresponding uniformly most powerful unbiased test (e.g., cf., \cite{Lehmann1986}). 
\end{example}

\subsection{A Unified Notion of $p$-Value for Univariate $\theta$}

Based on the concepts of CD supports, we are ready to provide a unified notion of $p$-value as follows.
\begin{definition}
Let  $\Theta_0 \subseteq \Theta$ and $\Theta_0 = \bigcup_{i=1}^K \Theta_{0i}$, where $\Theta_{0i} \in \mathcal{A} ~(i=1,\cdots,k)$ are disjointed. 
\begin{equation} \label{p:ev1}
p(\bfX_n, \Theta_0)=\max_{1\leq i \leq k} S^+_n(\Theta_{0i}),
\end{equation}
where $S^+_n$ is the full support defined in $(\ref{full:support})$ under CD $H_n(\theta)$. 
\end{definition}

Note that, when $\Theta_0 \in \mathcal{A}$, $p(\bfX_n, \Theta_0) \equiv S^+_n(\Theta_{0})$. In such cases, the validity of $p(\bfX_n, \Theta_0)$ as a $p$-value mapping can be shown based on the validity of $S^+_n(\Theta_{0})$.

\noindent
{\bf The validity of $p({\bf X}_n, \Theta_0)$ where $\Theta_0$ is a union of intervals --}
In practice, there is an increasing demand of non-standard types of hypothesis testing, where the null spaces are not restricted in $\mathcal{A}$ (\ref{interval_type}).
For instance, in a bio-equivalence problem, the parameter of interest $\theta$ is a measurement for assessing bio-equivalence of two formulations of the same drug or two drugs, e.g., $\theta=\mu_1-\mu_2$ or $\theta=\frac{\mu_1}{\mu_2}$, where $\mu_1$ and $\mu_2$ are the population means of bioavailability measures of the two formulations/drugs. Let $\theta_l$ and $\theta_u$ be some known bio-equivalence limits (e.g., $\theta_u=1.25,\theta_l=0.8$), the following testing problem often considered:
$H_0: \theta \in (-\infty, \theta_l] \cup [\theta_u,\infty) ~versus~ H_A: \theta \in (\theta_l, \theta_u)$.
More generally, the intersection-union test \citep{Berger:1982} has the following form:
\be \label{test_IU}
H_0: \theta \in \bigcup_{i=1}^K \Theta_{0i} ~~{versus}~~ H_A: \theta \in \bigcap_{i=1}^K \{ \Theta \setminus \Theta_{0i} \},
\ee
where $\Theta_{0i}$'s are disjointed intervals.   
The validity of the $p$-value mapping $(\ref{p:ev1})$ can be shown by the following theorem. 

\begin{theorem} \label{max_pvalue}
Suppose that, for $i=1,\cdots,K$, $p_i$ is the corresponding (limiting) $p$-value of the statement $\mathcal{H}_{0i}: \theta \in \Theta_{0i}$, $\Theta_{0i} \in \mathcal{A}$. Then, $p = \max \{ p_i; i=1,\cdots,K \}$ is the (limiting) $p$-value of the statement $\mathcal{H}_{0}: \theta \in \bigcup_{i=1}^K \Theta_{0i}$. 
\end{theorem}

To measure the evidence to the null space $\Theta_{0}=\bigcup_{i=1}^K \Theta_{0i}$, we turn to consider $K$ hypothesis testing problems corresponding to null space $\Theta_{0i}$, $i=1,\cdots,K$. For each $\mathcal{H}_{0i}: \theta \in \Theta_{0i}$, $p_i=S^+_n(\Theta_{0i})$ provides the full support and a $p$-value. Then, $p(\bfX_n, \Theta_0)=\max_{1\leq i \leq K} p_i$ measures the largest evidence among $\Theta_{0i}$'s. For any $\alpha \in (0,1)$, $p_i<\alpha$, for all $i=1,\cdots,K$, implies $p(\bfX_n, \Theta_0)<\alpha$. On the one hand, the proper design of each $p_i$ guarantees the size of test $\mathcal{H}_{0i}$, then the size of test $\mathcal{H}_0$ is guaranteed; on the other hand, in order to reject $\mathcal{H}_0$, we need to reject all $\mathcal{H}_{0i}$. In sum, the evidence to $\Theta_0$ is small only if the evidence to every $\Theta_{0i}$ is small, and the evidence to $\Theta_0$ is large if the evidence to some $\Theta_{0i}$ is large. 
The idea of handling a bio-equivalence test by ``two one-sided tests" \citep{Schuirmann:1981} can be considered as a special case. 
A simple and clear way of calculating $p$-value for bio-equivalence test is given in the following example. 

\begin{example}[continues=eg1] 
Consider a bio-equivalence test $H_0:\theta\in (-\infty,a]\cup[b,\infty)$ vs. $H_A:\theta \in (a,b)$, where $a$, $b$ are known. 
We can obtain 
$
p({\bfy}_n, \Theta_0) = \max\{S_{n}^D((-\infty, a]) , S_{n}^D([b, \infty))\}
=  \max\{ \Phi(\sqrt{n}(a-{\bar y}_n)/s_n) , 1 - \Phi(\sqrt{n}(b-{\bar y}_n)/s_n)] \}. 
$
\end{example}

In addition, based on Theorem \ref{max_pvalue}, where $\Theta_{0i} \in \mathcal{A}$ ($i=1,\cdots,k$), $p({\bfX}_n, \Theta_0)$ allows us to provide $p$-values more broadly than  the regular intersection-union test (\ref{test_IU}). For instance, the results still hold when some or even all $\Theta_{0i}$'s are singletons.

\subsection{Guidelines of constructing $p$-value mappings}

Up to this point, a unified and comprehensive notion of $p$-value is provided.
It is important to notice that $p$-value construction is not unique and modification might be available in case-by-case scenarios (e.g., see the discussions in Remark 1).
Whatever the way of $p$-value construction, the bottom line is that the defined mapping $p(\cdot,\cdot)$ satisfies the two requirements in the performance-based {\it Definition \ref{pval_def_map}} ; i.e., i) for all $\theta \in \Theta_0$, 
$p(\bfX_n, \Theta_0)$ is stochastically equal to or larger than Uniform$[0,1]$ (at least asymptotically); ii) for all $\theta \in \Theta \setminus {\Theta}_0$,
$p(\bfX_n, \Theta_0)$ goes to 0, as $n \go \infty$.
Correspondingly, almost all modifications in the field of hypothesis testing problems, not restricted in the $p$-value approaches, are concerning two key points:
a)  guarantee the size of the test, especially when the sample size $n$ is limited;
b) achieve better power. 

As to the comparison of two $p$-value mappings under the same scenario, we emphasize that there is a trade-of between controlling the size of the test and pursuing better power. 
For each individual case or application, this trade-off should be best determined by domain experts.
For example, consider the case with $\Theta_0=[a,b]$, since $S^D_n(\Theta_0) \leq S^+_n(\Theta_0)$, for the same asymptotic size, testing by $S^D_n(\Theta_0)$ will reject more, and therefore have better power.
However, for small samples, $S^D_n(\Theta_0)$ may not guarantee the size and may be overly aggressive compared to $S^+_n(\Theta_0)$. 
The limitations of $S_{n}^D(\Theta_0)$ have been discussed in Remark 1.
If controlling the size is often of the primary concern, we may consider $S^+_n(\Theta_0)$ as a modification of $S_{n}^D(\Theta_0)$.
First, $S^+_n(\Theta_0)$ provides a comprehensive and unified notion of $p$-value (for any $\Theta_0 \in \mathcal{A}$). 
Second, when $\Theta_0$ is narrow and $n$ is not large, 
$S^+_n(\Theta_0)$ is preferred in terms of controlling the size of the test, since $S^+_n(\Theta_0) > S_{n}^D(\Theta_0)$ (c.f., Figure \ref{fig.h1} \& \ref{fig.h0} in the simulation study).


\begin{remark} \label{max_of_support}
In the context of intersection-union test (\ref{test_IU}), there exists another recommendation of the $p$-value mapping as $\max_{1 \leq i \leq k}\{ S_n^D(\Theta_{0i}) \}$  (cf. e.g. \cite{Sin:Xie:Str:2007}). When all $\Theta_{0i}$'s are intervals, the validity is obvious based on Theorem 3. Since $p(\bfX_n, \Theta_0) \geq \max\{ S_n^D(\Theta_{0i}); i=1,\cdots,k \}$, testing by the latter will have better power. However, it can be overly aggressive in controlling the Type I error, especially when one or more $\Theta_{0i}$ is singleton or small interval. 
Moreover, when all $\Theta_{0i}$ are singletons, $\max_{1 \leq i \leq k}\{ S_n^D(\Theta_{0i}) \}$ is not applicable and $p(\bfX_n,\Theta_0)$ should be used. 
\end{remark}

\begin{remark} \label{power_pval}
To enhance power, we may consider the following $p$-value mapping construction. 
Let  $\Theta_0 \subseteq \Theta$ and $\Theta_0 = \bigcup_{i=1}^K \Theta_{0i}$, where $\Theta_{0i} \in \mathcal{A} ~(i=1,\cdots,k)$ are disjointed. Write the piecewise $p$-value $p_i=S_n^{+}(\Theta_{0i})$, and the corresponding ordered $p$-values $p_{(i)}$, $i=1,\cdots, K$. Consider
\begin{eqnarray}
p^*(\bfX_n, \Theta_0)=p_{(1)}-p_{(2)},
\end{eqnarray}
where $p_{(1)}$ is the largest one among $p_i$'s and $p_{(2)}$ is the second largest one. 
\begin{example}[continues=eg1] 
Consider a bio-equivalence test $H_0:\theta\in (-\infty,a]\cup[b,\infty)$ vs. $H_A:\theta \in (a,b)$, where $a$, $b$ are known. 
We can obtain 
$
p^*({\bfy}_n, \Theta_0) = |S_{n}^D((-\infty, a]) - S_{n}^D([b, \infty))|.
$
\end{example}
\end{remark}

\begin{remark}
The aforementioned CD approaches of constructing $p$-value mappings in the intersection-union test for a single parameter can also accommodate more complex settings such as some multi-parameter cases (as seen in the example in section 4 of \cite{Berger:1982}). 
More specifically, consider $\theta={\theta_1 \choose \theta_2}$ and $H_0:  \in \{ a_1 \leq \theta_1 \leq b_1\} \cup \{ a_2 \leq \theta_2 \leq  b_2\}$, $p(\bfX_n, \Theta_0)$ can still be applied. 

While noting that the CD is used as a general tool to formulate and interpret $p$-values, CD itself does not rely on the null or alternative hypothesis. 
Therefore, the ``supports'' of multiple (mutually exclusive) sets under CD can be obtained identically as in a univariate case.
And once the CD is obtained, the proposed CD-based mappings can derive $p$-values for various choices of the null space $\Theta_0$.  
Potentially, this can answer the common complaint on the classical testing approach (as articulated in \cite{Marden2000}) that ``model selection is difficult''. 
\end{remark}

\section{CD-based Notions of $p$-Value for Multivariate Parameters} 

Our CD approach can also be used for multi-dimensional problems.
In this section, we extend our CD-based $p$-value mappings to multivariate hypothesis testings. 
With the help of bootstrap method and data depth, we can even skip the specification of CD, and build the $p$-value mappings by the bootstrap estimates directly. 
For the direct support, we simple consider the fraction of bootstrap samples that lie in the null space. 
For the indirect support, we apply the concept of data depth to determine the fraction of possible values in the parameter space, that are more outlying (less consistent) than the null space. 
In the following, we first give a brief description of a well-known notion of data depth, {\it Liu's Simplicial Depth} \citep{Liu:1990}.

Given $Z_1, \cdots, Z_m$ from the distribution $\Phi$ in $\mathcal{R}^k$, the {\it Simplicial Depth} of a given point $w \in \mathcal{R}^k$ with respect to $\Phi$ and the data cloud $\{ Z_1, \cdots, Z_m \}$ is 
\ba
D(\Phi; w) = P_{\Phi} \{ w \in S(Z_1, \cdots, Z_{k+1}) \},
\ea
where $S(Z_1, \cdots, Z_{k+1})$ is the closed simple whose vertices $Z_1, \cdots, Z_{k+1}$ are $(k+1)$ random observations from $\Phi$. The sample version of $D(\Phi; w)$ is $D(\Phi_m; w)$, where $\Phi_m$ denotes the empirical distribution of the sample $Z_i$'s. 
A data depth can be used to measure the ``depth'' or ``outlyingness'' of a given multivariate sample with respect to its underlying distribution, leading to a natural center-outward ordering of the sample points.

Let ${\bf X}_1, \cdots, {\bf X}_n$ be a random sample from $F$, a $d$-dimensional distribution, $d \geq 1$, and let ${\bf \theta}_F$ be a $k$-dimensional functional of $F$. Consider the testing problem 
${\mathcal H}_0: {\bftheta}_F \in {\bf \Theta}_0  \leftrightarrow {\mathcal H}_A: {\bftheta}_F \notin {\bf \Theta}_0$,
where ${\bf\Theta}_0$ is a point or a region. 
 Let ${\bftheta}_n^* \equiv {\bftheta}_n^*({\bf X}_1^*, \cdots, {\bf X}_n^*)$ be a bootstrap estimate of ${\bftheta}_F$, where ${\bf X}_1^*, \cdots, {\bf X}_n^*$ is a bootstrap sample drawn with replacement from ${\bf X}_1, \cdots, {\bf X}_n$. 
Denote the sampling distribution of ${\bftheta}_n^*$ as $G_n^*$, and the data depth as $D(\cdot ; \cdot)$. Then, based on the full support (\ref{full:support}), we propose 
\be \label{p:ev3}
p_n( {\bf \Theta}_0) = P^*({\bftheta}_n^* \in {\bf \Theta}_0) + P_{G_n^*}( {\bftheta}_n^*: D(G_n^*; {\bftheta}_n^*) \leq \inf_{{\bftheta}_0 \in {\bf \Theta}_0}  D(G_n^*; {\bftheta}_0) ).
\ee

Equation (\ref{p:ev3}) can be viewed as a fine-tuning of the notions of $LP$s proposed by \cite{Liu:Singh:1997}. 
On the one hand, the direct measurement of the space ${\bf \Theta}_0$ is the empirical strength probability (ESP) that ${\bftheta}_n^* \in {\bf \Theta}_0$. 
On the other hand, to tackle the problem where $\bfTheta_0$ is a small region or even a point, we consider the indirect evidence based on the bootstrap estimates having fewer depth than any point in ${\bf \Theta_0}$. 
When ${\bf \Theta}_0= \{ {\bftheta}_0 \}$ is a point, ${p}_n( \{ {\bftheta}_0 \} )=P_{G_n^*}\left\{ {\bftheta}_n^*: D(G_n^*; {\bftheta}_n^*) \leq D(G_n^*; {\bftheta}_0)\right\}$, which is the limiting $p$-value of in $\{ {\bftheta}_0 \}$ 


\begin{remark}
In (\ref{p:ev3}), we generalize $S_{n}^{IND}$ (\ref{indirect:support}) (defined for univariate cases) to a measure of the strength of evidence about ``how likely'' we get possible $\theta$ values that are at least as outlying (inconsistent) as any $\theta_0 \in \bfTheta_0$.
Thus, instead of two-tailed intervals in the univariate cases, the tailed regions are taken into account. Generally, the extended indirect support can be written as
\be \label{extended_ind}
{S}_{n}^{IND*}( \bfTheta_0 ) = S_{n}^D \left\{ \theta: H_{n}(\theta) \leq \inf_{\theta_0 \in \bfTheta_0} H_{n}(\theta_0) \right\}. 
\ee
In univariate cases, ${S}_{n}^{IND*}( \Theta_0 )$ is equivalent to ${S}_{n}^{IND}( \Theta_0 )$, when the CD has a unimodal and symmetric density function.
\end{remark}


To justify that (\ref{p:ev3}) is a $LP$, one needs to show ${p}_{n}(\bfTheta_0)$ converges weakly to ${Uniform}[0,1]$ in distribution.
Let $\bftheta_n \equiv \bftheta_n(\bfX_1, \cdots, \bfX_n)$ be a estimate of $\bftheta_F$ and $\forall$ $\bftheta_0 \in \bfTheta_0$.
$L_n$ denotes the distribution of $a_n(\bftheta_n-\bftheta_0)$ for a positive sequence $\{ a_n \}$ satisfying that $a_n \go \infty$ as $n \go \infty$. 
Then, we say $L_n \go L$, D-regularly, if as $n \go \infty$,
i) $L_n$ converges weakly to $L$; 
ii) $\sup |D(L_n; w)-D(L;w)| \go 0$. 

\begin{theorem} \label{Thm_multi}
Let $\bfTheta_0$ be a closed-connected region.  
For a boundary point $\bftheta_0$ in $\bfTheta_0$, where $\bfTheta_0$ admits a unique tangent plane. 
Assume that $L_n \go L$, D-regularly, $L_n \go L$, D regularly a.s.. Here, $L$ is a continuous distribution symmetric around $0$.
Let $L$ be the c.d.f. of the random variable $T$, assume that distribution of $D(L;T)$ is continuous. 
Then, ${p}_{n}(\bfTheta_0)$ converges weakly to ${Uniform}[0,1]$ as $n \go \infty$.
\end{theorem}

Here, since the indirect evidence goes to 0 as $n \go \infty$ and the direct evidence is proven to converge weakly to Uniform [0,1]  (see \cite{Liu:Singh:1997}), the proof is omitted.   
In practice, we would calculate $m$ values of $\bftheta_n^*$, say $\bftheta_{n,1}^*, \cdots, \bftheta_{n,m}^*$, the empirical distribution of which is denoted as $G_{n,m}^*$. 
Then, (\ref{p:ev3}) can be obtained as a combination of the following two parts: 
[i] the fraction of $\bftheta_{n,i}^*$'s that are in $\bfTheta_0$, i.e. $m^{-1} \sum_{i=1}^m I\{ \bftheta_{n,1}^* \in \bfTheta_0 \}$; 
[ii] the fraction of $\bftheta_{n,i}^*$'s  that are not in $\bfTheta_0$ have less depth than any $\bftheta_{n,i}^*$ in $\bfTheta_0$, namely, $m^{-1} \sum_{i=1}^m I\{ D(G_{n,m}^*;\bftheta_{n,i}^*) < \min_{\bftheta_{n,j}^* \in \bfTheta_0} D(G_{n,m}^*;\bftheta_{n,j}^*) \}$. 
Here, $I\{ \cdot \}$ is an indicator function with $I\{ A \}=1$ if $A$ occurs and $I\{ A \}=0$ otherwise. 

\begin{remark}
For multivariate cases, the shape of $\bfTheta_0$ can be various and even irregular. To guarantee the size of a test using the $p$-value, we may require a safe way of constructing a $p$-value mapping. 
Consider the situation where $\bfTheta_0$ has finite boundary points, which form a set ${\partial {\bfTheta}_0}={\overline {\bfTheta}_0} \setminus {\bfTheta_0}^o$, where ${\overline {\bfTheta}_0}$ is the closure of $\bfTheta_0$ and $\bfTheta_0^o$ is the interior. We consider another $p$-value mapping as
\be \label{p_max}
p^{m}({\bfX}_n,\Theta_0) = \max\{ S_{n}^D(\Theta_0) , \max_{v \in {\partial \Theta_0}}{S}_{n}^{IND}(v) \}.
\ee
The validity can be proven based on Lemma \ref{lemma_combination}. 
Here, the indirect support for some representative boundary points are listed separately. By reporting the largest value among these supports (direct support and those separate indirect supports), we are choosing a measure with the strongest strength, among all the evidence in support of $H_0$. 
Note that $p^{m}({\bfX}_n,\Theta_0)$ is generally safe in controlling Type-I error, but can be quite conservative in some cases.  

Moreover, under the settings of (\ref{p:ev3}), if $\bf\Theta_0$ has finitely many non-smooth boundary points, say ${\bf\theta}_1, {\bf\theta}_2, \cdots, {\bf\theta}_m \in \partial {\bf\Theta}_0$, we can construct 
\be \label{pval_max_multi}
p^{m}_n({\bfTheta}_0) = \max\{ p_n( {\bf\Theta}_0), p_n( \{ {\bf\theta}_1 \}), \cdots, p_n( \{ {\bf\theta}_m \}) \},
\ee
the form of which is similar to the $p$-value mapping (\ref{p:ev1}).  
We use CD supports to measure the strength of evidence to certain part of $\bf\Theta_0$ from different angle (direct or indirect, region or point), and then, we maximize the evidence by taking the largest value among the piecewise supports. 
In some cases, this approach can be effective to control the size of a test (c.f., Figure 5 \& 6 in the simulation study).
\end{remark}

\section{Illustrative Examples}
\subsection{Simulation Studies}
The simulation work, which has two parts, is designed to show that our CD approach is easily adaptable to various forms of the null space. As to some unusual ones, the solutions of existing testing methods have not been available or cannot be easily obtained. 
{\bf Part I} involves testing on the one-dimensional normal mean, while {\bf Part II} tackles with bivariate cases, of which the $p$-value mappings are constructed based on bootstrap and data depth as shown in Section 4. 
In addition, the simulation studies show the great simplicity of our CD approach. 
Especially, when specific form of CD is used and the CD density function belongs to common distributions (like in {\bf Part I}), our $p$-values can be obtained merely by simple integrals over intervals. 

{\bf Part I: testing the univariate normal mean.}
In this part, we show that our proposed mapping $p(\bfX_n, \Theta_0)$ in (\ref{p:ev1}) are broadly effective for tests where $\Theta_0$ is an interval (see Fig. \ref{fig.h1}), including standard one- and two- sided test; and some nonstandard cases where $\Theta_0$ is a union of disjointed small intervals (see Fig. \ref{fig.h2}). 

For simplicity, all results are simulated from standard normal variable $X$. The parameter of interest $\theta$ is the mean of $X$, and the true value is zero. 
In each procedure, a sample of size $n$ (30, 200, etc.) is taken, and the CD of $\theta$ can be obtained as illustrated in {\it Example 3}.
For each given $n$ and $\Theta_0$, this procedure is repeated 50 times, and the QQ-plots of the quantiles of the simulated $p$-values against theoretical quantiles based on Uniform[0,1] are provided.

We first consider the cases where $\Theta_0$'s are all (generalized) intervals and contain the true value zero:
(1a) $\{0\}$; (1b) $[-0.01,0.01]$; (1c) $[-0.5,0.5]$; (1d) $[0,0.1]$; (1e) $[0,1]$; (1f) $[0,\infty)$. The sample size $n$ is chosen as $30$ or $200$.
The corresponding QQ-plots are given in Figure \ref{fig.h1}. 
The empirical distributions of our proposed $p(\bfX_n,\Theta_0)$ are close to Uniform[0,1] (especially when $n=200$), not only for standard tests with one-sided (1a) and two-sided (1f) null space, but also for nonstandard tests with interval null spaces (1b),(1d) and (1e). Here, we have uniform under (1a), (1d), (1e) and (1f), simply because the true parameter is $\Theta_0$ itself or on the boundary of it. Under (1b), although the interval is narrow and the direct support is almost zero, the indirect support dominates and $p(\bfX_n,\Theta_0)$ is close to uniform. The only exception from uniform is (1c) where $p(\bfX_n,\Theta_0)$ is stochastically larger than Uniform[0,1] and tend to cluster around one when $n$ increases to 200, because the true value of $\theta$ is in the interior of $\Theta_0$ and $\Theta_0$ is not narrow. In sum, when $p(\bfX_n,\Theta_0)$ is used as a $p$-value in a test with an interval-type $\Theta_0$, it can guarantee the size of the test. 
 
\begin{figure}[h!]
\begin{center}
\includegraphics[height=9cm, angle=0]{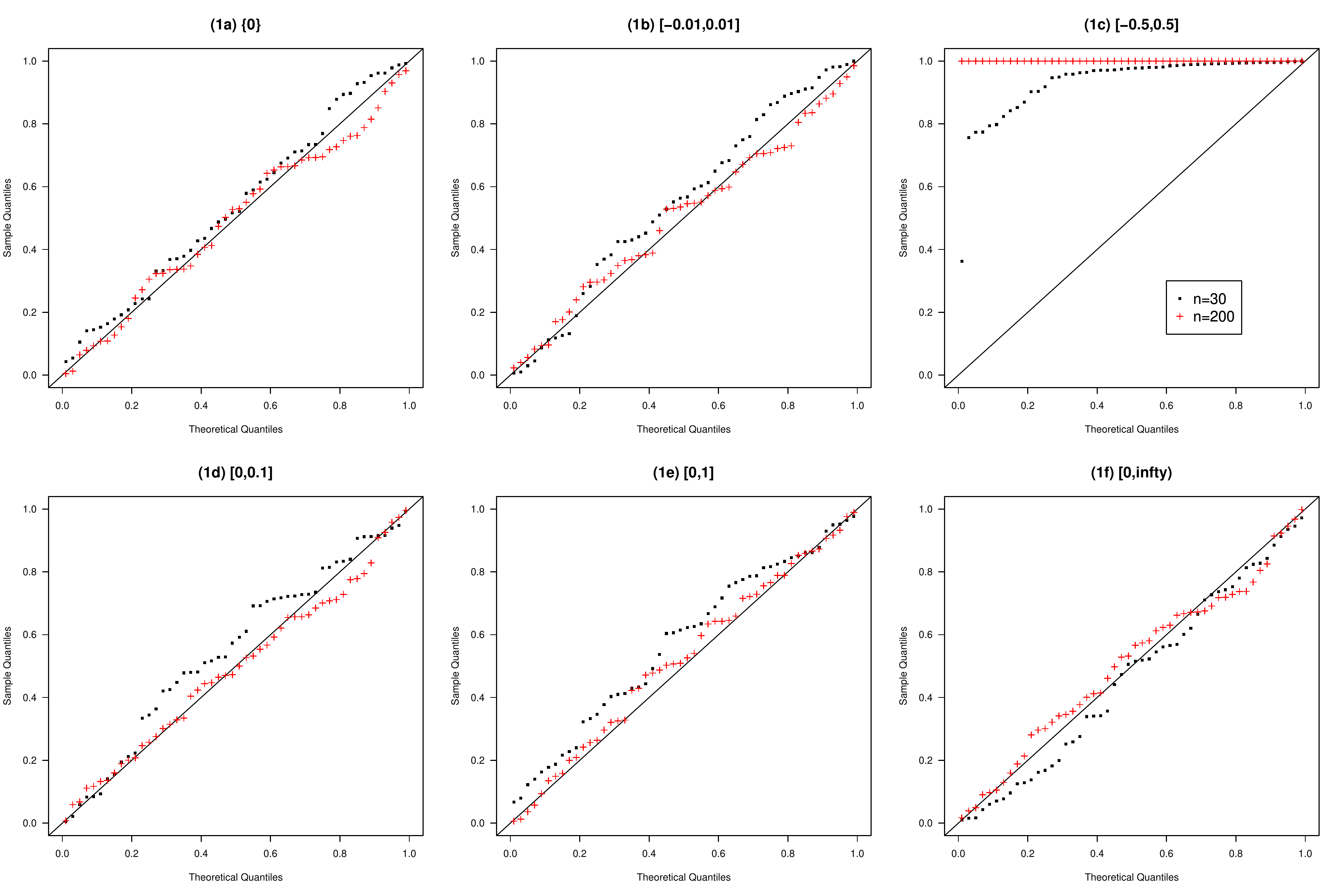}  
\end{center}
\caption{QQ-plots of quantiles of $p(\bfX_n,\Theta_0)$ values defined in (\ref{p:ev1}) against uniform quantiles. The true mean is contained in the null space $\Theta_0$: (1a) $\{0\}$; (1b) $[-0.01,0.01]$; (1c) $[-0.5,0.5]$; (1d) $[0,0.1]$; (1e) $[0,1]$; (1f) $[0,\infty)$. The sample size $n$ is taken as 30 (black dot) or 200 (red cross). }
\label{fig.h1}
\end{figure} 
 
In addition, to compare the performance of $p(\bfX_n,\Theta_0)$ with the direct support $S_n^D(\Theta_0)$, the QQ-plots of $S_n^D(\Theta_0)$ for (1b) \& (1d) are given in Figure \ref{fig.h0}. 
It is shown that, even when $n$ is as large as $5000$, the size of the test can be hardly controlled. 
Although $S_n^D(\Theta_0)$ is asymptotically valid to be used as a $p$-value mapping and can generally provide better power compared to $p(\bfX_n,\Theta_0)$, when $\Theta_0$ is very narrow and $n$ is limited, $S_n^D(\Theta_0)$ as an integral over $\Theta_0$ is often small, which possibly leads to over rejection of correct null.  
This shortcoming can be mitigated by our proposal, compared with (1b) \& (1d) in Figure \ref{fig.h1}, showing the benefits of involving the concept of indirect support to measure the evidence. 
This phenomenon implies that the width of interval has an undesired but non-negligible effect on making inference on $\theta$, and our proposal can be a useful solution. 
Note that $S_n^D(\Theta_0)$ is equivalent to the posterior probability of $\Theta_0$ under flat prior, thus, this ``narrow interval'' issue also exists in Bayesian inference.

\begin{figure}[h!]
\begin{center}
\includegraphics[height=5.5cm, angle=0]{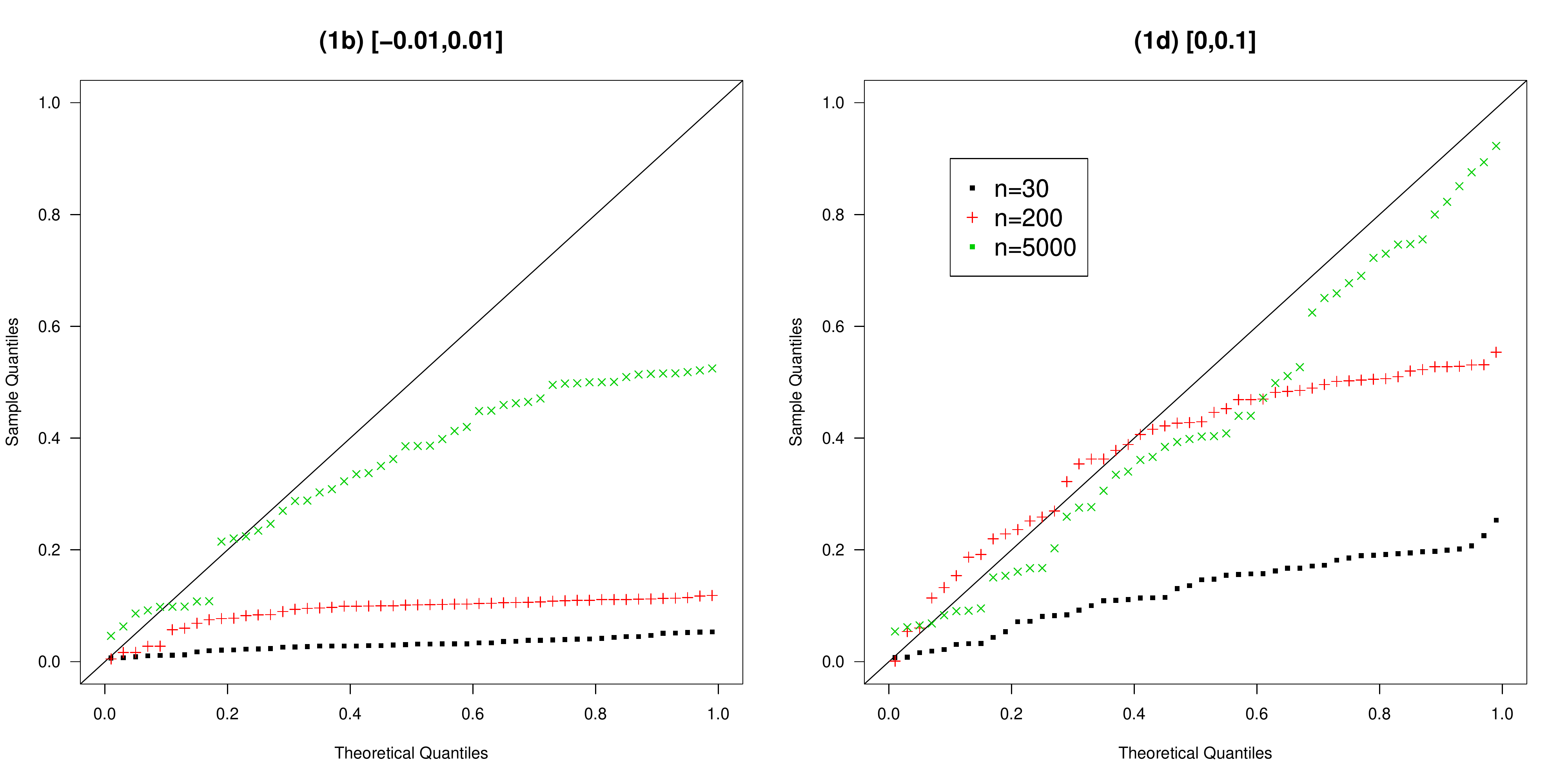}  
\end{center}
\caption{QQ-plots of quantiles of $S_n^D(\Theta_0)$ values against uniform quantiles. The true mean is contained in the null space $\Theta_0$: (1b) $[-0.01,0.01]$; (1d) $[0,0.1]$. The sample size $n$ is taken as 30 (black), 200 (red) and 5,000 (green). }
\label{fig.h0}
\end{figure}

Furthermore, we consider tests where $\Theta_0$'s have union forms: (2a) $[-\infty,0] \cup [0.5,+\infty]$; (2b) $[-0.04,-0.03] \cup [-0.01,0.01] \cup [0.02,0.03]$; (2c) $[0,0.1] \cup [0.5,0.6] \cup [1,1.1]$. 
In such cases, the approach of testing using rejection regions is quite complicated, since the construction of rejection regions relies on the choice of significance level $\alpha$ and the choices of rejection regions can be arbitrary. Note that, as an inference tool and before making any decision, our $p$-value approach does not depend on $\alpha$.
$p(\bfx_n, \Theta_0)$ defined in (\ref{p:ev1}) calculates the $p$-value for each interval first, and then reports the largest one among them. 
The true value is still included in the null spaces.  
Figure \ref{fig.h2} gives the QQ-plots. 
For all the four cases, $p(\bfX_n,\Theta_0)$ is close to Uniform[0,1]. Thus, the size of a test using $p(\bfx_n, \Theta_0)$ as a $p$-value is guaranteed. 
 
In comparison with $p(\bfX_n,\Theta_0)$, $S_n^D(\Theta_0)$ is also considered and the results of case (2b) \& (2c) are reported in Figure \ref{fig.h3}. When $\Theta_0$ is constructed by small intervals, like in (1b) \& (1d), the performance of $S_n^D(\Theta_0)$ is interfered, even for relatively large $n$. 
Moreover, when $\Theta_0$ only contains some singletons, $S_n^D(\Theta_0)$ is not applicable. 
The proposed $p(\bfX_n,\Theta_0)$ instead, does not have these issues. 
Again, if the CD we are using is the Bayesian posterior, such issues cannot be ignored for calculating the posterior probability of $\Theta_0$..
 
\begin{figure}[h!]
\begin{center}
\includegraphics[height=9cm,angle=0]{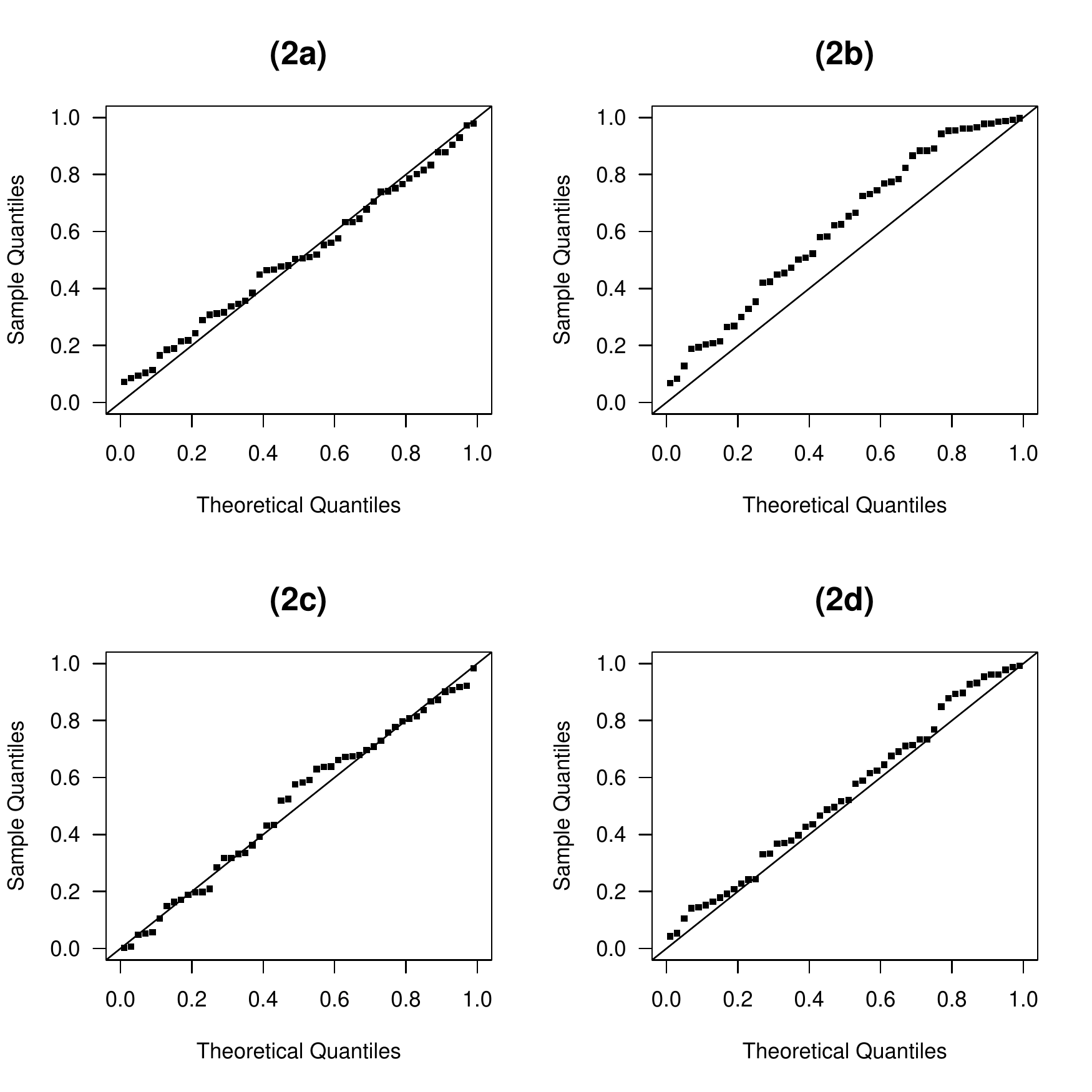}  
\end{center}
\caption{QQ-plots of quantiles of $p(\bfX_n,\Theta_0)$ values against uniform quantiles. The true mean is contained in the null space $\Theta_0$: (2a) $[-\infty,0] \cup [0.5,+\infty]$; (2b) $[-0.04,-0.03] \cup [-0.01,0.01] \cup [0.02,0.03]$; (2c) $[0,0.1] \cup [0.5,0.6] \cup [1,1.1]$; (2d) $\{ 0 \} \cup \{ 1 \}$. The sample size $n$ is taken as 30 (black dot).}
\label{fig.h2}
\end{figure}

  \begin{figure}[h!]
\begin{center}
\includegraphics[height=5.5cm,angle=0]{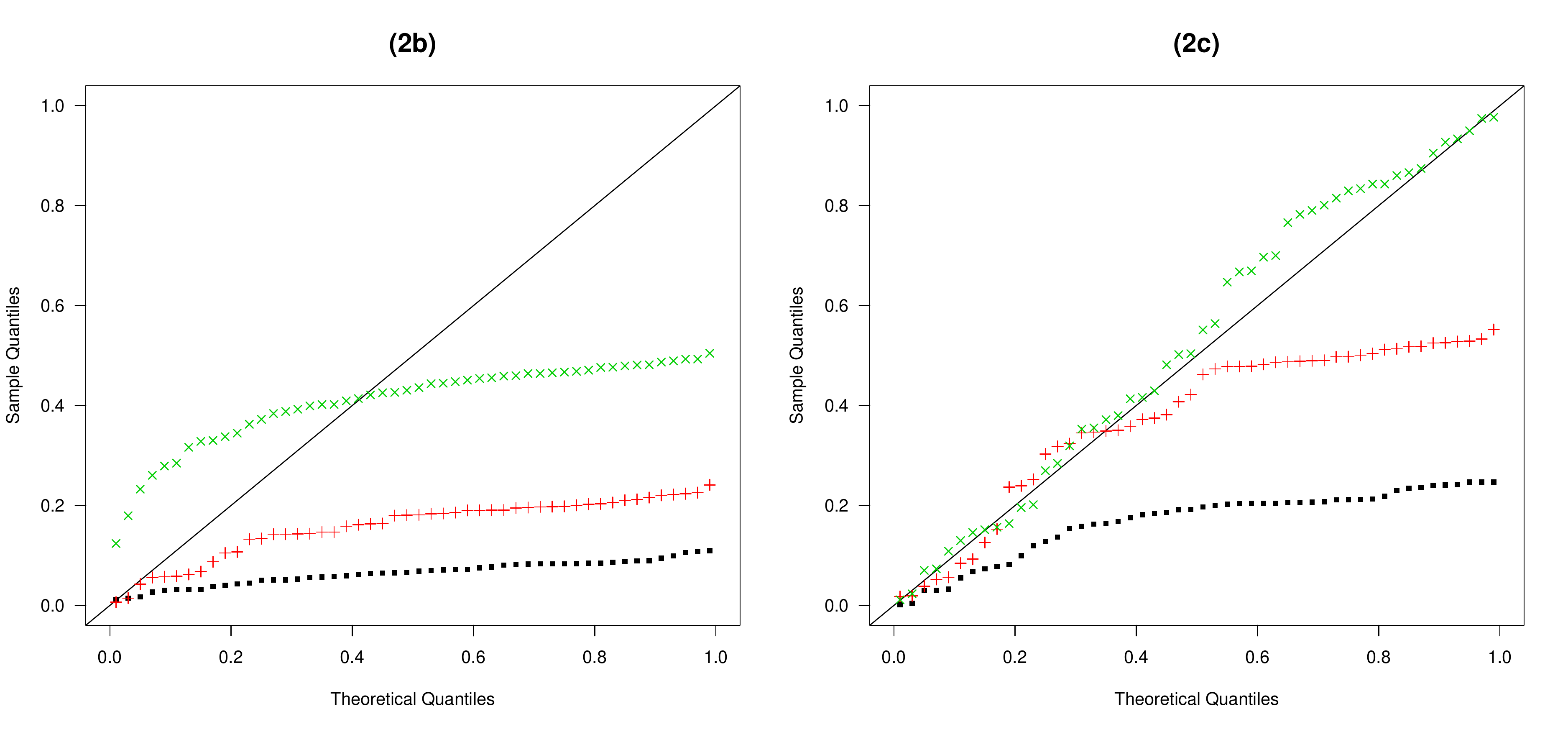}  
\end{center}
\caption{QQ-plots of quantiles of $S_n^D(\Theta_0)$ values against uniform quantiles. The true mean is contained in the null space $\Theta_0$: (2b) $[-0.04,-0.03] \cup [-0.01,0.01] \cup [0.02,0.03]$; (2c) $[0,0.1] \cup [0.5,0.6] \cup [1,1.1]$. The sample size $n$ is taken as 30 (black dots), 200 (red cross) and 2000 (green).}
\label{fig.h3}
\end{figure}
 


{\bf Part II: testing on the bivariate normal mean vector.}
The second part of the simulation study is used to demonstrate the applications of $p_n(\bf{\Theta}_0)$ proposed in (\ref{p:ev3}) by illustrating on some settings that involve nonstandard and complex null space existing in the literature \citep{Liu:Singh:1993}. 
Specifically, consider the hypothesis testing problem on the bivariate normal mean. Data are generated from a bivariate normal $(X_1, X_2)$, with mean $(0,0)$ and the covariance matrix 
\tiny
$\left( \begin{array}{cc}
1 & 0.8 \\
0.8 & 4
\end{array} \right)$.
\normalsize
The parameter of interest is the mean vector $\bftheta$. As a multi-dimensional case, the developments are based on bootstrap and the simplicial depth \citep{Liu:1990, Liu:Singh:1993}. 

For each given $\bf{\Theta}_0$ under $H_0: \bftheta \in \bf{\Theta}_0$, a sample of size $n$ (30 or 200) is taken from $(X_1,X_2)$, and for each sample $500$ bootstrap samples are drawn to compute $500$ bootstrap estimates of the mean vector. 
The proposed $p$-value mapping $p_n(\bf{\Theta}_0)$ is the ESP (the fraction of the bootstrapped estimates falling inside $\bf{\Theta}_0$), plus the fraction of bootstrap estimates outside $\bf{\Theta}_0$ that have lower data depth than those inside $\bf{\Theta}_0$. 
Different choices of $\bf{\Theta}_0$ are considered for illustrations:
\begin{enumerate}[(a)]
\item Rectangles with corners $\{ (-1,-1),(1,-1),(1,1),(-1,1) \}$; 
\item Rectangles with corners $\{ (-1,-4),(0,-4),(0,4),(-1,4) \}$; 
\item the complement of the quadrant $\{ (x_1,x_2): x_1 >0, x_2>0 \}$; 
\item Rectangles with corners $\{ (0,0),(0,-4),(-1,-4),(-1,0) \}$; 
\item Rectangles with corners $\{ (-0.1,-0.1),(0.1,-0.1),(0.1,0.1),(-0.1,0.1) \}$.
\end{enumerate}
For each given $\Theta_0$, the above procedure is repeated $50$ times to obtain $50$ values of $p_n(\bf{\Theta}_0)$, to form the corresponding QQ-plots. 
The results are shown in Figure \ref{fig.h5}. 

\begin{figure}[h!]
\begin{center}
\includegraphics[height=9cm,angle=0]{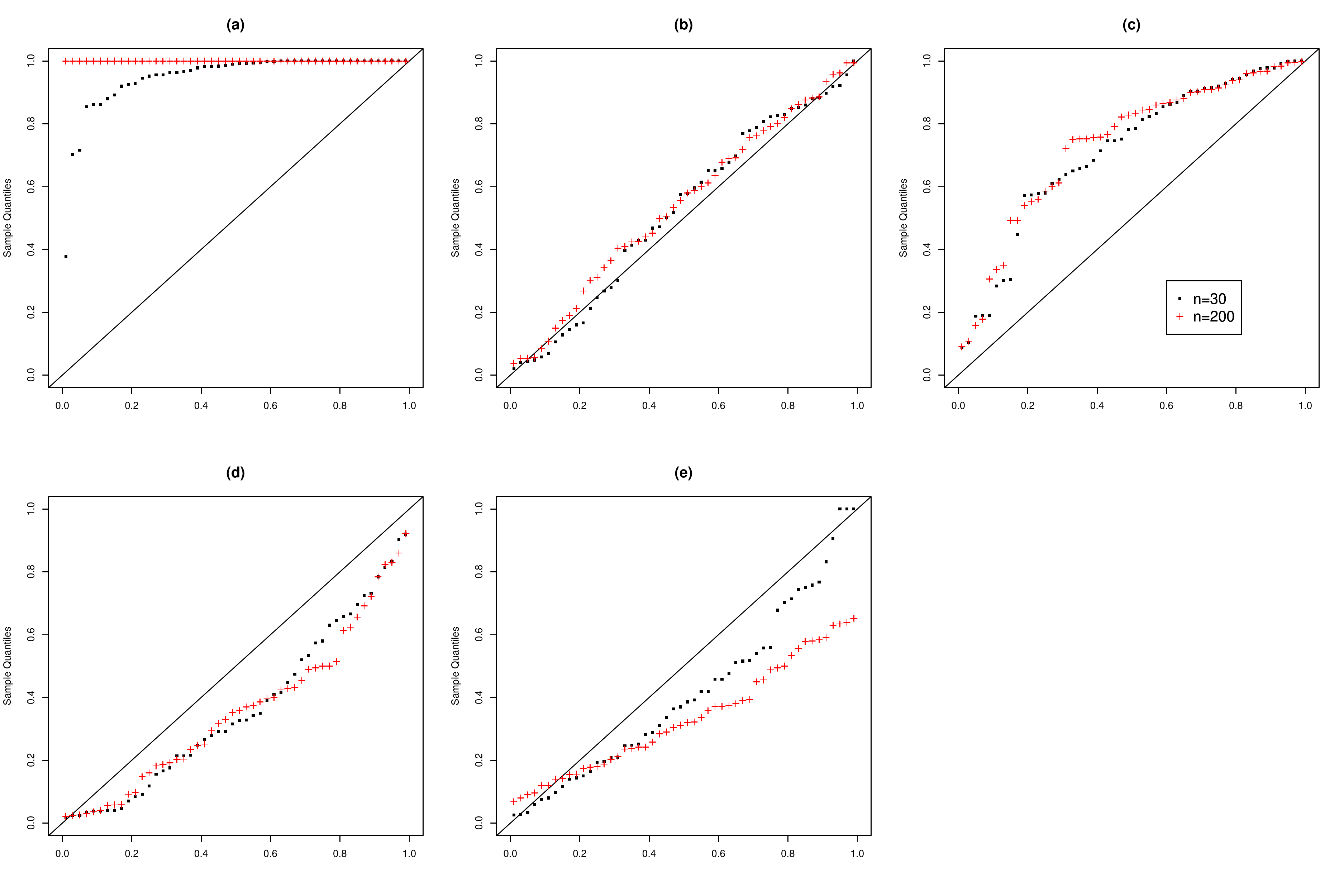}  
\end{center}
\caption{QQ-plots of quantiles of $p_n(\bf{\Theta}_0)$ values against uniform quantiles in Simulation Part II. The sample size $n$ is taken as 30 (black dots) or 200 (red cross).}
\label{fig.h5}
\end{figure}

For (a), since the true parameter value is in the interior of $\Theta_0$, $p_n(\bf{\Theta}_0)$ tend to assume values close to 1. 
For (b), the QQ-plot indicates Uniform[0,1], because the true value is on the smooth boundary. 
For (c), $p_n(\bf{\Theta}_0)$ is shown to be stochastically larger than Uniform[0,1], because the true value is a boundary point around a concave region. 
Up to (c), $p_n(\bf{\Theta}_0)$ works fine to guarantee the size of any test using it as a $p$-value. 
For (d) \& (e), 
$p_n(\bf{\Theta}_0)$ are clearly stochastically smaller than Uniform[0,1], even if $n$ is increasing from 30 to 200, leading to possibly over rejection of correct null space. 
Although the performance of $p_n(\bf{\Theta}_0)$ can be better than using ESP alone, we further apply $p^{m}_n(\bf{\Theta}_0)$ in (\ref{pval_max_multi}) to obtain improvements as shown in Figure \ref{fig.h6}.  
Here, in the QQ-plots, $p^{m}_n$ values are clear above the 45 degree line, which ensures that the test outcome be conservative (controlling the Type I error of the test).

\begin{figure}[h!]
\begin{center}
\includegraphics[height=4.5cm,angle=0]{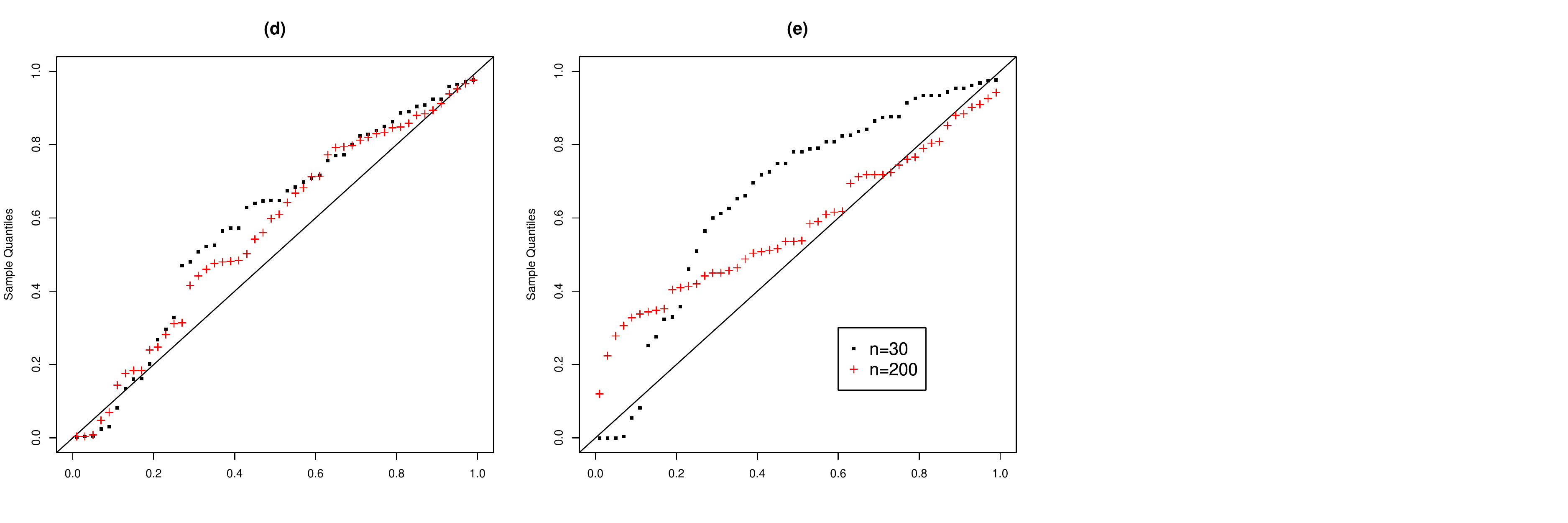}  
\end{center}
\caption{QQ-plots of quantiles of $p^{m}_n(\bf{\Theta}_0)$ values against uniform quantiles in Simulation Part II. The sample size $n$ is taken as 30 (black dots) or 200 (red cross).}
\label{fig.h6}
\end{figure}


\subsection{Real data applications}

\subsubsection{\textsf{Many $\alpha$ values, but only one-time calculation of $p$-value}} \mbox{}

In this section, a study on bio-equivalence problem presented to FDA is discussed. 
Almost all existing testing methods on this problem use the rejection region approach, of which the construction relies on the specified choice of the significance level ($\alpha$). 
However, in many areas such as political science, medical science, psychology and so on, different treatment actions may be employed to different levels of $\alpha$. 
For example,  \cite{Benjamin2017} suggests that the $p$-value threshold for making discoveries should be moved from 0.05 to 0.005. 
Therefore, it is likely that in the future FDA may adapt different levels of significance ($\alpha$ values) for different actions. 
We illustrate in this example that the proposed $p$-value is ideally suited for this situation, in that the $p$-value is presenting the strength of the evidence from data in support of the null. 
In such cases, a study of $p$-value can be used to make immediate decision while rejection region approach requires to redo the calculation procedure for different possible $\alpha$ values. 

\textsf{ Application I: the bio-equivalence problem.} Consider a two-period, crossover designed bio-equivalence study provided in \cite{Chow:Liu:2008} (example 3.6.1 on page 70).
The objective is to compare test (T) and reference (R) formulations of a drug product.
The study was conducted with 24 healthy volunteers. During each dosing period, each subject was administered either five 50 mg tablets (T formulation) or 5 mL of an oral suspension (R formulation). Blood samples were obtained at 0 hour before dosing and at various times after dosing. AUC values from 0 to 32 hours were calculated using the trapezoidal method.
Based on preliminary tests, there are no period and carryover effects.
Let {$\mu_T$, $\mu_R$} be the population means of AUC from T and R.
The bio-equivalence test of interest is 
\begin{equation} \label{realdata1}
\mathcal{H}_0: \theta \in (-\infty, \theta_l] \cup [\theta_u, \infty) {~versus~} \mathcal{H}_A: \theta \in (\theta_l, \theta_u). 
\end{equation}
where {$\theta=\mu_T-\mu_R$}.
Here, the limits $\theta_l$ and $\theta_u$ of the interval are predetermined. 
The bio-equivalence test based on Schuirmann's two one-sided t-tests becomes the standard approach in bio-equivalence studies, c.f., e.g. \cite{Hartmann:etal:1995, Berger:Hsu:1996, Kutta:etal:1999}.
We apply our $p$-value approach as follows. 
Note that, although the FDA bio-equivalence guidelines suggests $0.05$ as $\alpha$, there have been criticisms of $p$-values about using the magic number $0.05$. 

Denote the least square means (direct sample means for T and R) of the test and reference formulation as $\bar{Y}_T$, $\bar{Y}_R$, respectively;  and the pooled sample standard deviation of paired difference as $\hat{\sigma}_d^2$. Based on a central student {\it t} distribution with ($n_1+n_2-2$) degrees of freedom, a CD of $\theta$ can be constructed as 
$
H_{TR}(\theta)=F_{t_{n_1+n_2-1}}\left( \frac{\theta - (\bar{Y}_T-\bar{Y}_R)}{\hat{\sigma}_d \sqrt{\frac{1}{n_1}+\frac{1}{n_2}}} \right).
$
By the ``$\pm 20$ rule'', the bio-equivalence limits are chosen as $-\theta_l=\theta_u=16.51$.
Based on a sample data presented in the book, 
$\bar{Y}_T=80.272$, $\bar{Y}_R=82.559$,  $\hat{\sigma}_d^2=83.623$. For $H_0: \theta \in (-\infty,-16.51]\cup[16.51,\infty)$, the $p$-value by (\ref{p:ev1}) is calculated as $p(\bfX_n, \Theta_0)=\max\left\{\int_{(-\infty,-16.51]} dH_{TR}(u),{\int_{[16.51,\infty)} dH_{TR}(u)}\right\}=0.000479$.
The bio-equivalence of the test and reference formulations can be claimed based on this small $p$-value. 
\cite{Feng:etal:2006} also applied this idea of reporting the maximized one-sided $p$-values to a real-data example, but neither theoretical nor practical interpretation is provided. 




\subsubsection{ \textsf{nonstandard complex null space}} \mbox{}

We have shown in simulation study that our approach is easily adaptable to various forms of null space. 
In this section, we consider two real applications involving bivariate hypothesis testing problems with nonstandard complex null space, e.g., a rectangle. 
For such cases, we note that not only the rejection region approach requires a certain choice of $\alpha$ as illustrated in Section 5.2.1, but also the construction of rejection (or acceptance) region is often hard. 
For example, consider bivariate case with a rectangle null space, the acceptance region constructed by the test statistic in Hotelling's $T^2$-test is an ellipse, which cannot match the rectangular shape of the null. 
By our $p$-value approach, we can measure the strength of evidence in support of the null space directly, regardless of $\alpha$ or the form/shape of the null space.

{\textsf{Application II; Validation of simulation models.}}
Simulation models are often used to solve problems and to aid in making decisions. 
In the development of a simulation model, one important step is to determine whether it is an accurate representation of the system being studied \cite{Balci:Sargent:1981}.
The concern about whether this model is correct, is addressed through model validation. 
This validity is often tested under an {\it acceptable range of accuracy}, which refers to the acceptable agreement between the simulation model and the system under a given experimental frame. 
As a hypothesis testing problem, the null hypothesis can be generally stated as follows: model is valid for the acceptable range of accuracy under the set of experimental conditions. 
Specifically, the objective of this example provided by  \cite{Balci:Sargent:1982} is to determine whether the model represents a single server queueing system, which has two response variables, i.e. the average queue length for the first 500 customers ($X_1$) and the average waiting time in the system for the first 500 customers ($X_2$). The null hypothesis is specified as
$$
\mathcal{H}_0: |\mu_1^d| \leq 0.154,  |\mu_2^d| \leq 0.28, 
$$  
where $\mu_i^d$ is the population mean of the differences between the paired observations on the response $X_i$ from the model and system, $i=1,2$.
Since the null space is a rectangle, the conventional hypothesis testing procedures require modification, e.g. \cite{Balci:Sargent:1982, Sargent:2015, Sargent:Goldsman:Yaacoub:2015}. 
The existing methods rely on the construction of rejection (or acceptance) regions based on Hotelling's $T^2$-test under arbitrary choices of the Type I \& II errors ($\alpha$ and $\beta$), but the ellipse shape of acceptance region can never match the rectangular null, which leads to the fact that neither $\alpha$ nor $\beta$ can be achieved exactly.
In the following, we apply our $p$-value approach, which does not require specified $\alpha$ or $\beta$. 
The resulting $p$-values measure the strength of evidence in support of the null space. 

The sample data of size $15$ is provided in Table \ref{table.data}. An estimate of the variance-covariance matrix of differences between the paired observations on the model and system response variables was given as 
\tiny
$\left( \begin{array}{cc}
0.2162 & 0.4147 \\
0.4147 & 0.7959
\end{array} \right)$.
\normalsize
We consider the {\it Mahalanobis depth} \citep{Mahalanobis1936}, where the depth of a point $w$ relative to a data set $U_n$ is defined as
$
D(U_n;w) = \left[ 1+(w-\mu_U)' \Sigma_U^{-1} (w-\mu_U)  \right]^{-1}. 
$
Here, $\mu_U$ and $\Sigma_U$ are the mean vector and variance-covariance matrix. 
Based on (\ref{p:ev3}), the $p$-value can be obtained as $0.486$, which indicates that the differences between the model and system are acceptable.

\tiny
\begin{table}[h!]
\begin{center}
\begin{tabular}{ccccccccc}
\hline
Difference on $X_1$ & -0.255 & 0.201 & 0.008 & 0.014 & -0.146 & 0.321 & 0.097 & 0.679 \\
 & 0.361 & 0.269 & 0.153 & 0.329 & 0.283 & 0.657 & -0.314 \\
Difference on $X_2$ & -0.631 & 0.372 & -0.128 & 0.035 & -0.390 & 0.639 & 0.303 & 1.240 \\ & 0.398 & 0.505 & 0.207 & 0.465 & 0.438 & 0.905 & -0.458 \\
\hline
\end{tabular}
\end{center}
\caption[]{Selected sample data for validation of simulation model. }
\label{table.data}
\end{table}
\normalsize

\textsf{ Application III: Aircraft Landing Performance: Airbus versus Boeing.}
Consider two studies of different aircraft makes: Airbus 321 (A) and Boeing 737-400 (B). 
The objective is to compare the landing performances of Airbus and Boeing, and provide the Federal Aviation Administration (FAA) advisory directives with landing performance guidelines. 
The FAA is the oversight agency responsible for regulating air traffic and safety. 
The improved capacity and increased flow at airports to accommodate the rapid growth of air traffic in the United States have led the FAA to initiate many new research efforts in aviation safety. 
The key tasks are to investigate the aircraft landing performance pertaining to operational safety guidelines and possibly set new advisory directive on landing operations. 
In particular, \cite{Vanes2005} reported that the most frequently reported aircraft landing incidents are runway overruns, and there is a significant increase in overrun risk when an aircraft has long {\it landing distance}. 
Here, overrun means landing aircraft are unable to stop before the end of the runway, and landing distance refers to the distance from the beginning of the runway to the aircraft touchdown point. 

\begin{figure}[h!]
\begin{center}
\includegraphics[width=12cm,angle=0]{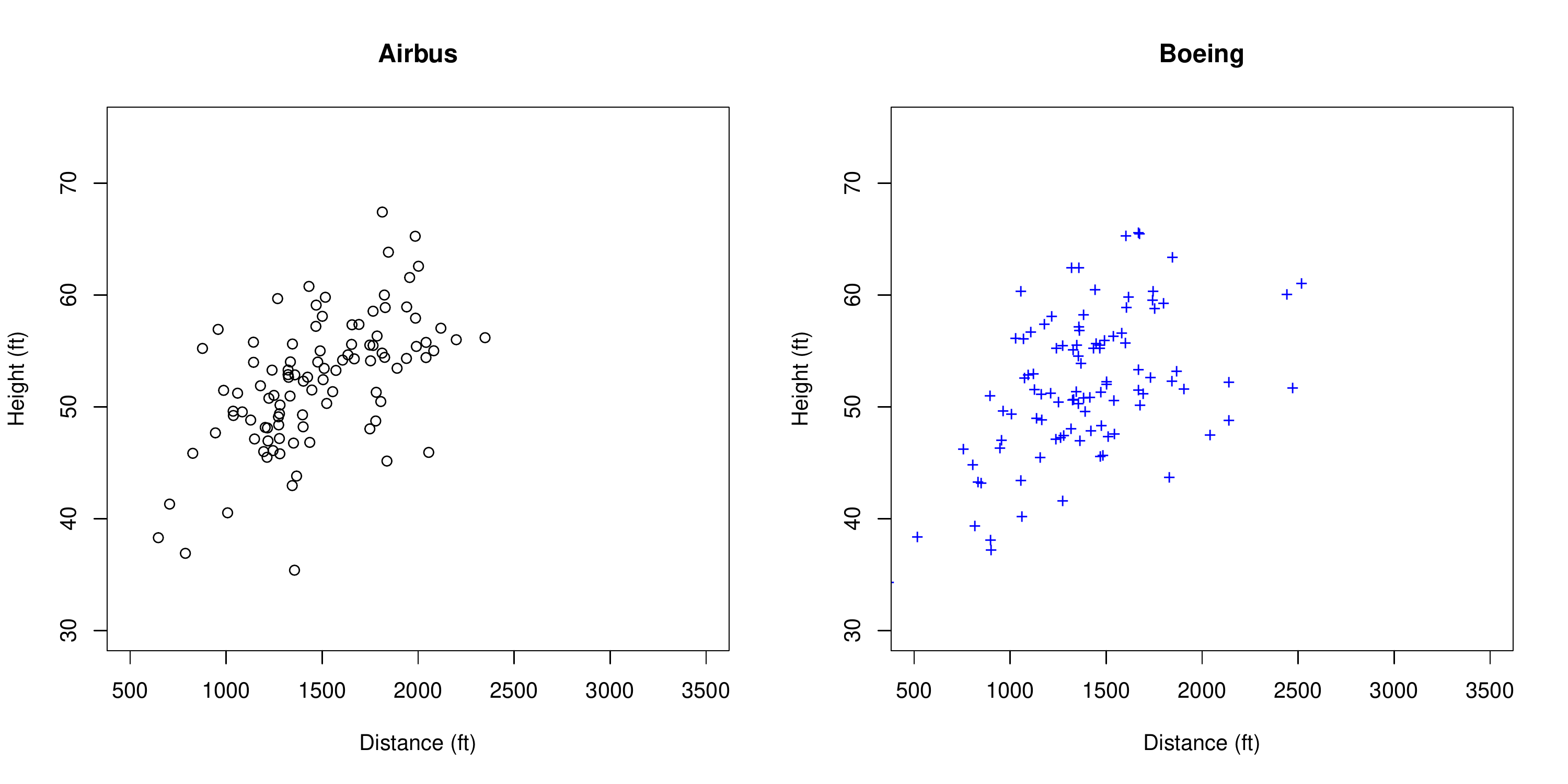}  
\end{center}
\caption{Sample data: Height and Distance for Airbus (left) and Boeing (right).}
\label{fig.app3}
\end{figure}

Specifically, we randomly selected $100$ samples from each type: Airbus and Boeing. A scatter plot is provided in Figure \ref{fig.app3}. 
The observation of each flight contains two variables: landing distance (in ft) and the height of airplane at the threshold (in ft). 
Correspondingly, denote the mean vector of Height and landing distance as $\bftheta=(\theta_1, \theta_2)$. 
For tests of hypotheses, the null hypothesis $\bfTheta_0$ we are interested in include:
(a) $\bfTheta_{01}= (1417.3,52)$; 
(b) $\bfTheta_{02}: 1000 \leq \theta_1 \leq 2500, 50 \leq \theta_2 \leq 55$; 
(c) $\bfTheta_{03}: 3000 \leq \theta_1 \leq 5000$;
(d) $\bfTheta_{04}: 1450 \leq \theta_1 \leq 1550, 51 \leq \theta_2 \leq 53$.
Similar to the example of validation of simulation models, almost all existing hypothesis procedures depend on the construction of rejection regions and require the set-up of significance level. Therefore, we apply our proposed $p$-value mapping (\ref{p:ev3}) to measure the strength of evidence obtained from data in support of $\bfTheta_0$. 
Here, (\ref{p:ev3}) is a unified notion of the $p$-value mapping for various $\bfTheta_0$, some of which is nonstandard (e.g. the small rectangle in (d)).

\begin{figure}[h!]
\begin{center}
\includegraphics[width=8cm,angle=0]{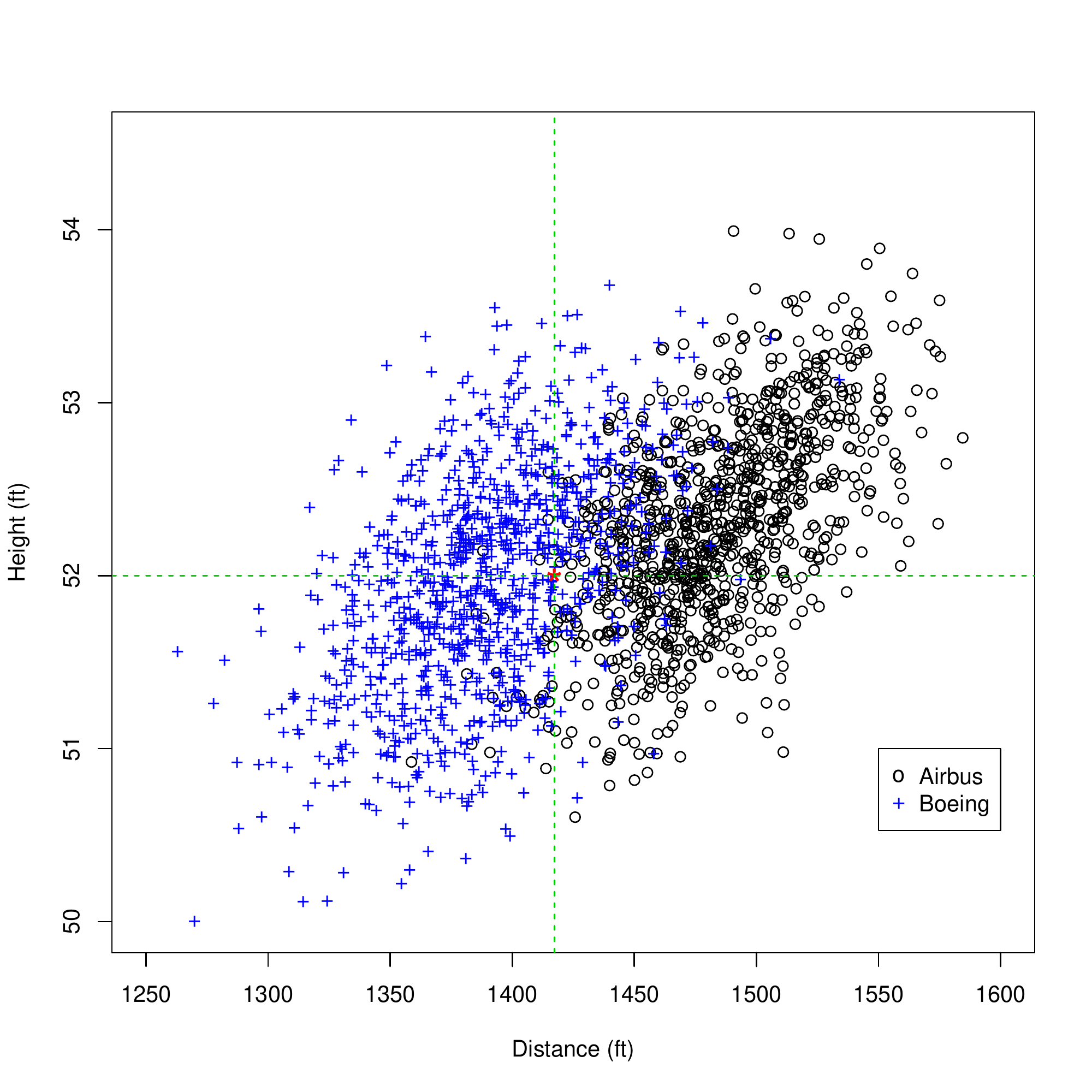}  
\end{center}
\caption{Bootstrap estimates for Airbus (black) versus Boeing (blue). $\Theta_0$ in (a): (1417.3,52) is marked as red star.}
\label{fig.app3.1}
\end{figure}

For each choice of the null space, we calculate $p$-values for both Airbus and Boeing. 
First, the point null in (a) is obtained by the overall mean of the whole dataset. 
Based on bootstrap method, we obtained $p_A=0.075$ and $p_B=0.671$, implying that $\bfTheta_{01}$ is a relatively reasonable advisory directive guideline for Boeing, but not Airbus. 
For (b), the rectangular parameter space (a ``safe landing" zone) is given by restricting the landing distance in [1000,2500] and height at threshold in [50,55]. 
For both Airbus and Boeing, $p$-values are close to $1$, indicating that the sample data support $\bfTheta_{02}$. 
In addition, (c) is the "advisory" values, meaning that if the distance is over $3000$, we need to care about the landing performance and detect the reason. By these samples, there is no "extreme" observations, and $p_A=p_B=0$ showing no supports of $\bfTheta_{03}$. 
Finally, we consider a small rectangle in (d). Based on bootstrap samples, we have $p_A=0.678$, $p_B=0.041$. Therefore, $\bfTheta_{04}$ can be used as a standard guideline of landing for Airbus, but not Boeing. 
In sum, the ``safe landing'' zone $\bfTheta_{02}$ and the ``advisory'' zone $\bfTheta_{03}$ are supported by the sample data. But we detect some difference in the landing performance of Airbus and Boeing aircrafts. For instance, $\bfTheta_{04}$ is only supported by Airbus sample data. Compared to Airbus, the landing distance of Boeing aircrafts is shorter. 


 
\section{Discussion}


In this article, we emphasize that $p$-value is evidential, and more specifically, it is an assessment of the strength of evidence coming from observed data. 
Closely connected to the logic ``proof by contradiction'', this idea is not new throughout the development of Fisher's significance test. 
For example, \cite{Hubbard:2004} pointed out ``the $p$-value from a significance test is regarded as a measure of the implausibility of the actual observations (as well as more extreme and unobserved ones) obtained in an experiment or other study, assuming a true null hypothesis''. 
However, this may cause confusions because large $p$-value in fact does not imply implausibility, while small $p$-value (e.g. $p$-value$<0.05$) is treated as ``evidence against'' $H_0$. 
The underlying reason is that there was no clear definition of $p$-value with evidence-based interpretations. Our proposed definition together with CD's construction fill this gap. 
$p$-value is indeed the ``evidence in support of'' $H_0$.
 This line of thinking also matches the Bayesian argument that for one-sided test, the posterior of probability of $\Theta_0$ with non-informative prior is typically equivalent to one-sided $p$-value. 
Therefore, under frequentist's point of view, we provide an evidence-based interpretation. Comparing two $p$-values under the same scenarios (say, $0.8$ and $0.5$), we can say $0.8$ indicates more support of the null than $0.5$. 

Our proposed definition of $p$-value highlights the two performance-based characteristics, directly following the key logic. 
It allows us to broaden the concept of $p$-value to a function that assesses the strength of evidence coming from observed data supporting the statement.
As to the construction of $p$-value mappings, the CD representations of $p$-values are provided for a broad class of problems. First, one-sided and two-sided $p$-values are unified by combining the direct and indirect supports in our proposals. The direct support measures , while the indirect evidence is from "the enemy of enemy", taking account of the position adjustment. More generally, the proposed $p$-value mapping can present a measurement of evidence to not only interval-type hypotheses, but also union hypotheses.  Furthermore, multi-dimensional parameter cases are tackled with the help of bootstrap and data depth. Our proposals can cover almost all common hypotheses testing problems in practice. 
For many non-standard cases, the solutions of existing methods have not been available or cannot be easily obtained.  




There has been recent suggestions to abandon the use of $p$-values. But we want to emphasize that, the $p$-value does have its appealing characteristics, and it is far too early to say we should replace the $p$-value with alternative testing tools (e.g., confidence interval (CI), Bayes-Factor (BF)).  
In fact, those alternative approaches are closely connected to the $p$-value and have their own problems. 

Specifically, the preference for CIs is quite common in medical, social, and other applied sciences (e.g., \cite{Fidler2004}).
However, when we use CIs, we are still on the page with $p$-value rather than abandoning it. 
In most applications, to achieve decision making with a preset $\alpha$, the $p$-value is used to compare with $\alpha$, while the CI is equivalent to an acceptance region with confidence level $1-\alpha$.
Often, there is a one-to-one mapping between a CI and a significance level $\alpha$ (threshold for $p$-value), therefore, the testing results from CI and $p$-value should be equivalent, under the same choice of $\alpha$. 
Our CD approach provides a natural connection between $p$-value and CI, because i) $p$-value is constructed by CD supports; ii) CD provides CIs of all levels.  
Moreover, as an inference tool, $p$-value mapping does not rely on $\alpha$, while CI depends on $\alpha$. 
In practice, the choice of $\alpha$ can be arbitrary and various. 
For different $\alpha$ values, our $p$-value approach does not need to redo the computation procedure, but CIs require case-by-case computation, similarly as the construction of rejection region in hypothesis testing problems. 

There have also been suggestions from Bayesian literature to replace the use of $p$-value by BF (e.g., \cite{Kass1995}).
Although, for one-sided tests, the $p$-value and the posterior probability of the null (under flat priors) are actually equivalent \citep{CasellaBerger1987};
for two-sided tests, BF is not always well-defined especially under improper noninformative priors as observed in \cite{Berg:Dela:1987, Rousseau:2006}.
Moreover, the computation of BF involving marginal likelihood over null/alternative parameter space can be complicated. 

Recall that CD is a distribution estimator, and that Bayesian posteriors can be considered as a special case of CD. 
We can apply our construction procedure in (\ref{p:ev1}) to construct $p$-value mapping based on posterior distribution as follows. 
Suppose that the posterior distribution of $\theta \in \Theta$ is $\pi(\theta \mid \bfx)$.
For $\Theta_0 \subset \Theta$, let $\Theta_0^L= \{ \theta': \pi(\theta'|\bfx) \leq \inf_{\theta \in \Theta_0}\pi(\theta|\bfx) \}$. 
We have {\it a Bayesian version of the $p$-value}:
$$
{p}_{\bfx}^{\pi}(\Theta_0)= P^{\pi}(\Theta_0 \mid \bfx) + P^{\pi}(\Theta_0^L \mid \bfx),  
$$
where $P^{\pi}$ denotes the posterior probability. 
Here, the expressions $P^{\pi}(\Theta_0 \mid \bfx)$ and $P^{\pi}(\Theta_0^L \mid \bfx)$ are analogue to the direct support in (\ref{direct:support}) and the indirect support in (\ref{extended_ind}), respectively. 
Similarly, this construction allows us to interpret ${p}_{\bfx}^{\pi}(\Theta_0)$ as a measure of the strength of evidence under the Bayesian framework. 
Note that this line of interpretation is different from those suggested in \cite{Gelman1996,Rubin1984} in Bayesian inference, which roughly speaking, interpret Bayesian $p$-value as sort of an average of the frequentist $p$-value function over the the domain based on the prior or posterior.  


\appendix
\appendixpage
\addappheadtotoc
\subsection*{Appendix A. Proof of Proposition \ref{Prop1} }
\begin{proof}
In this proof, we show that ${pval}_1$ satisfies (a) \& (b) in {\it Definition \ref{pval_def_map}}. Note that, under the condition that large value of $T(X)$ is against $H_0$ and $G_{T,\theta}$ exists, ${pval}_2$ is equivalent to ${pval}_1$.
 
Write ${pval}_1(\bfx_n)  =  \sup_{\theta \in \Theta_0} p_{\theta}(\bfx_n)$, where $p_{\theta}(\bfx_n) \equiv {P}_{\theta}\{T(\bfX_n) \geq T(\bfx_n)\}$.
For any $\theta \in \Theta_0$, define $F_S$ as the c.d.f. of $S=-T(\bfX_n)$, we have c.d.f.
$p_{\theta}(\bfx_n) = 1 - G_{T,\theta}\{T(\bfx_n)\} = F_S(s)$. 

For (a), define a random variable $Y=p_{\theta}(\bfX_n) = F_S(S)$ and $U_y \equiv \{ s : F_S(s) \leq y \}$. If $U_y$ is half-closed $(-\infty, s_y]$, 
$\Pr[Y \leq y] = \Pr[F_S(S) \leq y] = \Pr[S \in U_y] = F_S(s_y) \leq y$. 
If $U_y$ is half-open $(-\infty, s_y)$, by continuity of probability,
$\Pr[Y \leq y] = \Pr[F_S(S) \leq y] = \Pr[S \in U_y] = \lim_{s \go s_y} F_S(s) \leq y$. 
Thus, given any $\alpha \in (0,1)$, $P_{\theta}[p_{\theta}(\bfX_n) \leq \alpha] \equiv \Pr[Y \leq \alpha] \leq \alpha$. Since ${pval}_1(\bfX_n) \geq p_{\theta}(\bfX_n)$, we have $P_{\theta}\{{pval}_1(\bfX_n) \leq \alpha\} \leq \alpha$, which is true for any $\theta \in \Theta_0$. 

For (b), note that $\{{pval}_1(\bfx_n) \leq \alpha \} \supseteq \{p_\theta(\bfx_n) \leq \alpha \}=\{ G_{T,\theta}\{T(\bfx_n)\} \geq 1-\alpha \}$. Based on the construction of $T$, we have $P_{\theta'} (G_{T,\theta}\{T(\bfX_n)\} \geq 1-\alpha) \go 1$, as $n \go \infty$, for any $\theta' \in \Theta \setminus \Theta_0$. Therefore, ${pval}_1$ satisfies (b). The results follow. 
\end{proof}

\subsection*{Appendix B. Proof of Lemma 1}
\begin{proof}
(a) Let $C=(-\infty,a]$. For any $\theta \leq a$, ${P}_{\theta} \left( H_{n}((-\infty,a]) \leq \alpha \right) \leq$ \\ ${P}_{\theta} \left( H_{n}((-\infty,\theta]) \leq \alpha \right)=\alpha$. $C=[b,\infty)$ can be dealt with in a similar way. 

(b) It suffices to show the argument with the sup over $\theta \in \Theta_{0j}$, $j=1,\cdots,k$. First, consider $\Theta_{0j}=(-\infty,a]$ and any $\delta >0$, on one hand,
$\sup_{\theta \in \Theta_{0j}} P_{\theta} \{ S_{n}^D(\Theta_0) \leq t \}
\geq P_{\theta=a} \{ S_{n}^D(\Theta_0) \leq t \}
\geq P_{\theta=a} \{ S_{n}^D(\Theta_{0j}) \leq t -\delta \} + o(1)
=t-\delta+o(1)$.
On the other hand, 
$\sup_{\theta \in \Theta_{0j}} P_{\theta} \{ S_{n}^D(\Theta_0) \leq t \}
\leq \sup_{\theta \in \Theta_{0j}} P_{\theta} \{ S_{n}^D(\Theta_{0j}) \leq t \}
=t.$
Thus, $\sup_{\theta \in \Theta_{0j}} P_{\theta} \{ S_{n}^D(\Theta_0) \leq t \} \go t$.  $\Theta_{0j}=[b,\infty)$ can be handled similarly. 
Second, for $\Theta_{0j}=[c,d]$, write it as the union of $\Theta_{0j1}=[c,\frac{c+d}{2}]$ and $\Theta_{0j2}=[\frac{c+d}{2},d]$.
Then, we can use the similar arguments as above to complete the proof. 
\end{proof}

\subsection*{Appendix C. Proof of Lemma 3}
\begin{proof}
Suppose that $p_1(\bfX_n, \Theta_0)$ is a $p$-value. 
First, since $p_1(\bfX_n, \Theta_0)$ satisfies (a) and $p_2(\bfX_n, \Theta_0) \geq p_1(\bfX_n, \Theta_0)$, $p_2(\bfX_n, \Theta_0)$ satisfies (a) as well. 
Second, $p_2(\bfX_n, \Theta_0)$ satisfies (b), because both $p_1(\bfX_n, \Theta_0)$ and $p_2(\bfX_n, \Theta_0)$ satisfy (b). 
Then, we can use the similar arguments as above to complete the proof when $p_1(\bfX_n, \Theta_0)$ is $LP$.
\end{proof}


\renewcommand{\baselinestretch}{1.2}
\tiny\normalsize 
\bibliographystyle{agsm}
\bibliography{Hypothesis.bib}



\end{document}